\documentclass{jfm}

\usepackage{graphicx}
\usepackage{natbib}
\usepackage{amsmath}
\usepackage{hyperref}
\usepackage{color}
\usepackage{comment}

\title[]{
Pressure Hessian and viscous contributions to velocity gradient statistics based on Gaussian random fields
}

\author[M. Wilczek and C. Meneveau]
{Michael Wilczek$^1$
  \thanks{Email address for correspondence: mwilczek@jhu.edu},
and Charles Meneveau$^1$}

\affiliation{$^1$Department of Mechanical Engineering \& Institute for Data Intensive Engineering and Science, The Johns Hopkins University, 3400 North Charles Street, Baltimore MD 21218}

\newcommand{\bs}{\boldsymbol}
\newcommand{\mat}{\mathsfbi}

\newcommand{\A}{\bs {\cal A}}
\renewcommand{\S}{\bs {\cal S}}
\newcommand{\W}{\bs {\cal W}}

\newcommand{\Aij}{{\cal A}_{ij}}
\newcommand{\Aik}{{\cal A}_{ik}}
\newcommand{\Akj}{{\cal A}_{kj}}
\newcommand{\Ajl}{{\cal A}_{jl}}
\newcommand{\Alk}{{\cal A}_{lk}}
\newcommand{\Akl}{{\cal A}_{kl}}
\newcommand{\Anp}{{\cal A}_{np}}
\newcommand{\Aii}{{\cal A}_{ii}}

\newcommand{\Aonekl}{{\cal A}_{1,kl}}
\newcommand{\Atwokl}{{\cal A}_{2,kl}}
\newcommand{\Aoneij}{{\cal A}_{1,ij}}
\newcommand{\Atwoij}{{\cal A}_{2,ij}}
\newcommand{\Aonemn}{{\cal A}_{1,mn}}
\newcommand{\Atwomn}{{\cal A}_{2,mn}}
\newcommand{\Atwonm}{{\cal A}_{2,nm}}
\newcommand{\Aoneik}{{\cal A}_{1,ik}}
\newcommand{\Aoneki}{{\cal A}_{1,ki}}
\newcommand{\Aonejl}{{\cal A}_{1,jl}}
\newcommand{\Aonelj}{{\cal A}_{1,lj}}

\begin{document}

\maketitle

\begin{abstract}
Understanding the non-local pressure contributions and viscous effects on the small-scale statistics remains one of the central challenges in the study of homogeneous isotropic turbulence. Here we address this issue by studying the impact of the pressure Hessian as well as viscous diffusion on the statistics of the velocity gradient tensor in the framework of an exact statistical evolution equation. This evolution equation shares similarities with earlier phenomenological models for the Lagrangian velocity gradient tensor evolution, yet constitutes the starting point for a systematic study of the unclosed pressure Hessian and viscous diffusion terms. Based on the assumption of incompressible Gaussian velocity fields, closed expressions are obtained as the results of an evaluation of the characteristic functionals. 
The benefits and shortcomings of this Gaussian closure are discussed, and a generalization is proposed based on results from direct numerical simulations. This enhanced Gaussian closure yields, for example, insights on how the pressure Hessian prevents the finite-time singularity induced by the local self-amplification and how its interaction with viscous effects leads to the characteristic strain skewness phenomenon.
\end{abstract}

\newpage

\section{Introduction}

The understanding of turbulence dynamics and statistics by theoretical means is hampered by two challenges: nonlinearity and non-locality. These challenges become especially evident when studying the smallest scales of turbulence in terms of the velocity gradient tensor $\mat A=\nabla {\bs u}$ (where ${\bs u}$ is the velocity vector).  The velocity gradient tensor gives a comprehensive characterization of the small scales. Its symmetric part, the rate-of-strain tensor, characterizes the local rates of deformation of fluid elements and, for example, determines the rate of kinetic energy dissipation. In addition to nonlinear advection, viscous diffusion and the interaction with the vorticity field, the rate-of-strain dynamics is also subject to non-local pressure effects through the pressure Hessian.  The isotropic part of the pressure Hessian preserves solenoidality of the velocity field, whereas the deviatoric part communicates information between distant points in the velocity gradient field.
The antisymmetric part of the velocity gradient tensor, the rate-of-rotation tensor, represents the vorticity and yields further insights into the small-scale coherent structures of the field. Also the vorticity is subject to nonlinear advection through the velocity field and viscous diffusion, but is locally stretched by the rate-of-strain tensor. 

With its wealth of information, the velocity gradient tensor has been subject to research for many decades. Whereas early works by \cite{betchov56jfm} revealed a number of important kinematic constraints on its statistical properties, the dynamical properties of the velocity gradient tensor moved into focus by the works of \cite{vieillefosse82jpf,vieillefosse84pha} and \cite{cantwell92pfa}. In these works, the local self-amplification of the velocity gradient tensor along Lagrangian fluid trajectories has been studied in detail neglecting the non-local pressure contributions and dissipative effects. This so-called restricted Euler approximation led to valuable insights into several features of small-scale turbulence. For example, these studies elucidated the preferential alignment of the vorticity vector with the principal axes of the rate-of-strain tensor, which is also observed in direct numerical simulations (DNS) studies \citep{ashurst87pof}. The shortcoming of this approximation is highlighted by the development of a singularity in finite time. This outcome reveals that pressure Hessian and viscous contributions have to be taken into account for a realistic, non-divergent statistical and dynamical description of the velocity gradient.  Later it has been shown by \cite{martin98pof} that the inclusion of a linear diffusion term prevents the singularity for initial conditions with moderate velocity gradient tensor values (as compared to the damping rate), while for larger values the quadratic nonlinearity still overpowers the linear damping.

These prior works revealed that the velocity gradient dynamics can be conveniently studied in a Lagrangian frame. This idea has been picked up and considerably extended by the works of \cite{chertkov99pof} and \cite{naso05pre}, who argued that a tetrad, whose corners are defined by four Lagrangian fluid particles, can be interpreted as a scale-dependent perceived velocity gradient tensor, giving insights into the statistical properties of not only the dissipative but also the inertial range of scales. The governing equations of motions have been closed by phenomenological arguments, where it has been assumed that the main effect of the pressure Hessian is to deplete the nonlinear self-amplification of the gradients.

Another related phenomenological approach, focusing on the smallest scales of fluid motion, has been proposed by \cite{chevillard06prl} who developed a closure based on the deformation of an infinitesimal fluid element. By taking into account only its recent history, closed expressions for the pressure Hessian and the viscous term have been obtained. While it has been found that this model yields a good description of many observed features, it has also been shown in a critical comparison with DNS data by \cite{chevillard08pof} that the model misses some features of the pressure Hessian. Further limitations concern the behaviour for large Reynolds numbers.  For a more extensive literature review including other phenomenological models, see \cite{meneveau11arf}.

The quality of the phenomenological models summarized here has reached a level where satisfactory agreement of numerical and experimental data is achieved for moderate Reynolds numbers. The complexity of both the closure problem itself and the already proposed models, however, renders progress based on further phenomenological refinements a challenge and motivates further theoretical investigations. Moreover, a theoretical justification of the proposed models remains elusive and motivates a deeper investigation of the pressure Hessian and viscous contributions. 

The current work aims at contributing to such investigations. To this end, we study the statistical properties of the velocity gradient tensor in terms of an exact, yet unclosed statistical evolution equation. This evolution equation describes the statistical properties of a class of fluid particles that share the same value of the velocity gradient tensor. The advantage of this approach is the rigorous formulation of the closure problem in terms of random fields, which serves as a starting point to establish well-controlled closures based on as few assumptions as possible. To obtain explicit expressions for the pressure Hessian and viscous terms, we then make the assumption that the velocity field can be represented by an incompressible Gaussian random field. This is known to be inaccurate due to, e.g., small-scale intermittency, but this approach allows for a fully analytical treatment. In particular, insights into the formal structure of the pressure Hessian and the viscous term are obtained without further ad-hoc assumptions. The merits and shortcomings of this Gaussian closure are discussed, and a generalization based on DNS observations is proposed. In this enhanced Gaussian closure, dimensionless parameters obtained from the analytical solutions are replaced by empirical fits that make use of DNS data. The dynamical features of this new closure illustrate how the non-local pressure Hessian contributions help to prevent a finite-time singularity and how the interaction with the diffusive ingredients of the dynamics allow for strain skewness and enstrophy production. Finally, predictions from this enhanced Gaussian closure are compared with DNS data.

\section{Velocity gradient dynamics and statistical description}

In this section we briefly review the basic equations of motion and then proceed to a detailed description of the statistical methods used in this paper.

\subsection{Velocity gradient tensor dynamics and non-local pressure contributions}

The dynamics of the velocity gradient $A_{ij}(\bs x,t)=\frac{\partial u_i}{\partial x_j}(\bs x,t)$ is obtained by taking the spatial gradient of the incompressible three-dimensional Navier-Stokes equation, yielding
\begin{equation}\label{eq:velgradevolution}
  \frac{\partial}{\partial t} \mat A + \bs u\cdot\nabla\mat A = -\mat A^2 - \mat H + \nu \Delta \mat A + \mat F  \, .
\end{equation}
Here $\bs u(\bs x,t)$ is the velocity field, $H_{ij}(\bs x,t)=\frac{\partial^2p}{\partial x_i \partial x_j}(\bs x,t)$ is the Hessian of the kinematic pressure field, $\nu$ is the kinematic viscosity, and $\mat F(\bs x,t)$ constitutes the gradient of a solenoidal large-scale forcing optionally included in the Navier-Stokes equation. This equation states that the velocity gradient tensor field is advected with the velocity field while being subject to self-amplification or self-attenuation, as well as pressure and viscous diffusion effects and optionally an external forcing. The influence of the self-amplification term is well understood thanks to its locality, and has been analysed extensively in the context of the restricted Euler model, for example, by \cite{cantwell92pfa}. The viscous term already incorporates some non-local information, because information from neighbouring points in the field is needed to evaluate the Laplacian. The pressure Hessian term is related to the velocity gradient tensor field by a Poisson equation,
\begin{equation}
  \Delta p = \mathrm{Tr} \left(  \mat H \right) = -\mathrm{Tr}\left( \mat A^2 \right) \, ,
\end{equation}
which is obtained from \eqref{eq:velgradevolution} by taking the trace and  using the fact that $\mathrm{Tr}\left( \mat A \right)=0$ and $\mathrm{Tr}\left( \mat F \right)=0$.
This shows that the isotropic part of the pressure Hessian is also local, such that the pressure Hessian can be decomposed into
\begin{equation}\label{eq:hessiandecomposition}
  \mat H = -\frac{1}{3}\mathrm{Tr}\left( \mat A^2 \right) \mat I + \widetilde{\mat H} \,
\end{equation}
where $\mat I$ is the identity matrix and $\widetilde{\mat H}$ is a traceless symmetric tensor containing all the non-local contributions of the pressure field. This can be made more explicit by the fact that the non-local pressure Hessian contribution is obtained from the velocity gradient field by a principal value integral \citep{ohkitani95pof} which (for unbounded flow) reads
\begin{equation}\label{eq:nonlocalpressurehessian}
  \widetilde{H}_{ij}(\bs x,t) = -\frac{1}{4\pi} \! \int_{\mathrm{P.V.}} \!\!\!\! \mathrm{d}\bs x' \left[ \frac{\delta_{ij}}{|\bs x-\bs x'|^3}-3\frac{(\bs x-\bs x')_i (\bs x-\bs x')_j}{|\bs x-\bs x'|^5}  \right] \mathrm{Tr}\left( \mat A(\bs x',t)^2 \right) \, .
\end{equation}
This relation stresses that the deviatoric part of the pressure Hessian contains highly non-local information from remote points in the fluid. Interestingly, its dependence on the velocity gradient tensor occurs only through a scalar invariant, $Q = -\frac{1}{2}\mathrm{Tr}\left(  \mat A^2 \right)$.

For the discussion later on it is also useful to decompose the velocity gradient dynamics into its symmetric and antisymmetric contributions, $\mat S = \frac{1}{2}\left( \mat A + \mat A^{\mathrm{T}} \right)$ and $\mat W = \frac{1}{2}\left( \mat A - \mat A^{\mathrm{T}} \right)$, respectively. In terms of these tensors (for the moment considering a flow without body force) the velocity gradient dynamics \eqref{eq:velgradevolution} takes the form
\begin{align}
  \frac{\partial}{\partial t} \mat W + \bs u\cdot\nabla \mat W &= -\mat S \mat W - \mat W \mat S + \nu \Delta \mat W \\
  \frac{\partial}{\partial t} \mat S + \bs u\cdot\nabla\mat S &= -\left[ \mat S^2 - \frac{1}{3}\mathrm{Tr}\left( \mat S^2 \right) \mat I \right] -  \left[ \mat W^2 - \frac{1}{3}\mathrm{Tr}\left( \mat W^2 \right) \mat I \right] -\widetilde{\mat H} + \nu \Delta \mat S  \, .
\end{align}
Alternatively, the antisymmetric part is readily expressed in terms of the vorticity vector, $\omega_i = -\varepsilon_{ijk}W_{jk}$, where $\varepsilon_{ijk}$ is the Levi-Civita tensor. Taking $\mat S$ and $\bs \omega$ as primary variables, the above equations can be written as
\begin{align}
  \frac{\partial}{\partial t} \bs \omega + \bs u\cdot\nabla \bs \omega &= \mat S \bs \omega +\nu \Delta \bs \omega \label{eq:vorticityeq}\\
  \frac{\partial}{\partial t} \mat S + \bs u\cdot\nabla\mat S &= -\left[ \mat S^2 - \frac{1}{3}\mathrm{Tr}\left( \mat S^2 \right) \mat I \right] -  \frac{1}{4}\left[ \bs \omega \bs \omega^{\mathrm{T}} - \frac{1}{3} \bs \omega^2 \mat I \right] -\widetilde{\mat H} + \nu \Delta \mat S  \, . \label{eq:rateofstraineq}
\end{align}
The velocity gradient dynamics becomes particularly clear from these equations when assuming $\mat S$ as initially diagonal without loss of generality.  According to \eqref{eq:vorticityeq}, vorticity is stretched or attenuated depending on the sign of the specific eigenvalue of the rate-of-strain tensor, in addition to advection and viscous diffusion. As can be seen from \eqref{eq:rateofstraineq}, the rate-of-strain is advected and diffuses. More importantly, however, it can be noted that the first term on the right-hand side is diagonal in the eigenframe of the rate-of-strain tensor and causes a self-amplification or self-attenuation of velocity gradients similar to that observed in Burgers dynamics. The second term also contributes to this amplification, but furthermore induces a rotation of the eigenframe depending on the vorticity vector. Understanding how the non-local pressure Hessian acts in this equation is one of the central topics in the present paper.

In the following we will explicitly consider a stochastic large-scale forcing. Under the assumption that the small-scale statistics of turbulence is independent of the large-scale forcing for sufficiently high Reynolds numbers, this particular choice does not appear to be a severe restriction, but it will turn out useful to make a connection to existing phenomenological models. Furthermore, it allows for a fully analytical treatment.

\subsection{Statistical evolution equation}

We now turn to a statistical description of the problem, focusing on $f(\A;t)$, the probability density function (PDF) of the velocity gradient tensor at a single point in the fluid. This function, which depends upon the nine  velocity gradient tensor elements and time,  contains rich information on the small-scale properties of turbulence, including the single-point statistics of vorticity, rate-of-strain and their mutual orientation.

To obtain an evolution equation for the probability function, we use the statistical framework of the Lundgren-Monin-Novikov hierarchy, which allows to derive exact, yet unclosed evolution equations for PDFs. A basic account on the methodology can be found in \cite{lundgren67pof,dopazo94trf,pope00book,wilczek11jfm} and \cite{friedrich12crp}. Appendix \ref{app:pdfeqderivation} gives a detailed derivation of the relations used in this section, but we here rather focus on a discussion of the physical implications of non-locality on the small-scale statistics of turbulence. The closure problem arising in this statistical framework can be cast in two different ways. One way is introducing conditional averages of, e.g., the pressure Hessian, with respect to the velocity gradient, the other way is to express the unclosed terms in terms of multipoint statistics. The two formulations give complementary perspectives on the problem, and in fact both perspectives turn out to be useful for the current work.

The PDF equation for the velocity gradient tensor in homogeneous turbulence reads (see also \cite{girimaji90pof})
\begin{align}\label{eq:pdfeq}
  \frac{\partial}{\partial t} f(\A;t) = &-\frac{\partial}{\partial \Aij} \left(\left[ -\left(\Aik\Akj-\frac{1}{3}\mathrm{Tr}\left( \A^2 \right) \delta_{ij}\right) - \big\langle \widetilde H_{ij} \big | \A \big\rangle + \big\langle \nu \Delta A_{ij} \big | \A \big\rangle \right] f(\A;t) \right) \nonumber \\
  &+ \frac{1}{2} Q_{ijkl}(\bs 0)\frac{\partial}{\partial \Aik}\frac{\partial}{\partial \Ajl} f(\A;t) \, ,
\end{align}
where $Q_{ijkl}(\bs 0)$ denotes the two-point covariance tensor of the gradients of the large-scale forcing evaluated at the origin. Without a stochastic forcing \eqref{eq:pdfeq} is a Liouville equation describing the conservation of probability. Due to the fact that the self-amplification and the isotropic part of the pressure Hessian are local, these terms appear closed. The only unclosed terms in this equation are the conditional averages of the non-local pressure Hessian and the viscous diffusion. The stochastic forcing, which only enters the equation through its covariance tensor, turns the PDF equation into a Fokker-Planck equation. The Fokker-Planck equation can be associated to a Langevin equation which establishes the connection to existing phenomenological models. This equation takes the form
\begin{equation}\label{eq:stochasticcharacteristic}
  \mathrm{d}\A = \left[ -\left( \A^2-\frac{1}{3}\mathrm{Tr}\left( \A^2 \right) \mat I \right) - \big\langle \widetilde{ \mat H} \big | \A \big\rangle + \big\langle \nu \Delta \mat A \big | \A \big\rangle \right] \mathrm{d}t + \mathrm{d}\mat F \, .
\end{equation}
Here, $\mathrm{d}\mat F$ is a stochastic forcing which is determined by the large-scale forcing applied to the velocity field (see appendix \ref{app:pdfeqderivationforcing} for more details). Again, the closure problem arises in terms of the conditional averages of the pressure Hessian and the viscous diffusion of the velocity gradient. This exact, yet unclosed equation has clear similarities with the dynamical phenomenological stochastic models for the Lagrangian velocity gradient evolution reported earlier \citep{girimaji90pof,Jeonggirimaji03,chevillard06prl,meneveau11arf}, which focus on modelling realizations of the process. The difference, however, is that this equation describes the statistical evolution of a \textit{class} of fluid particles which all share the identical value of the velocity gradient, whereas the Lagrangian velocity gradient models express the pressure Hessian and the diffusive term in terms of the velocity gradient tensor along that trajectory. 

The fact that the closure problem in the current formulation arises in terms of conditional averages allows us to study the unclosed terms based on general random fields. While this certainly does not solve the central problem, it will yield some interesting insights. Moreover, the assumption of Gaussian random fields   allows to  calculate the unclosed terms analytically. Both of these points will be discussed below in detail.

\subsection{Conditional viscous diffusion and conditional pressure Hessian}

To make contact with a formulation in terms of general random fields, we first have to establish the relation of the unclosed terms  to multipoint statistics. The unclosed terms in \eqref{eq:pdfeq} and \eqref{eq:stochasticcharacteristic} can be expressed in terms of two-point statistics, which underscores their non-local nature. For the following we will consider homogeneous isotropic turbulence. The technical background on obtaining the relations discussed in this paragraph are given in appendix \ref{app:relationtomultipoint}. 

The relation obtained for the viscous term reads
\begin{equation}\label{eq:conditionallaplaciantwopoint}
  \big\langle \nu \Delta_{\bs x_1} \mat A(\bs x_1,t) \big | \A_1 \big\rangle = \lim_{r \rightarrow 0}  \nu\Delta_{\bs r} \big\langle \mat A(\bs x_2,t) \big | \A_1 \big\rangle \, ,
\end{equation}
where we have introduced subscripts to discriminate the two points in space and the distance vector $\bs r=\bs x_2 -\bs x_1$. The viscous term contains non-local information in terms of the average velocity gradient at point $\bs x_2$ conditional on the value of the velocity gradient at point $\bs x_1$. Still, it can be considered as somewhat local because only the immediate neighbourhood of $\bs x_1$ has to be known to evaluate the derivative. 

For the conditional pressure Hessian we obtain
\begin{equation}\label{eq:conditionalpressurehessianmain}
\big\langle \widetilde H_{ij}(\bs x_1,t) \big | \A_1 \big\rangle = \frac{1}{2\pi} \! \int_{\mathrm{P.V.}} \!\!\!\!\, \mathrm{d}\bs r \left[ \frac{\delta_{ij}}{r^3}-3\frac{r_i r_j}{r^5}  \right] \big\langle Q(\bs x_2,t) \big | \A_1 \big\rangle \, .
\end{equation}
This result is interesting in two respects: First, the structure of the integral kernel assures that the conditional Hessian will be traceless and symmetric in $i$ and $j$, no matter what is assumed for $\big\langle Q(\bs x_2,t) \big | \A_1 \big\rangle$; second, this unclosed tensorial conditional average depends only on a conditional scalar expression, which is a considerable  simplification, especially for modelling. Still the conditional second invariant represents a rather complicated function, which may depend on all invariants that can be constructed from $\A_1$ and $\bs r$.

By expressing the conditional average in terms of two-point statistics, we have ``shifted" the closure problem from an unclosed expression for the joint single-point statistics involving the pressure Hessian and the velocity gradient to the joint two-point statistics involving the velocity gradient only. This allows for an evaluation in terms of general random fields, and we will pursue the special case of Gaussian random fields in the following sections.

We remark that up to now all of the discussed relations represent exact results. Once a more general theory of random fields is available, they can be used to obtain more elaborate closures.

\section{A closure based on Gaussian random fields}

Before presenting the central results of this paper, some words on why to choose a closure based on Gaussian random fields are in order, especially given the fact that it is generally known that the multipoint structure of the velocity field in turbulent flows exhibits important non-Gaussian features such as intermittency and skewness of velocity increments and gradients (see, e.g., \cite{frisch95book}). The justification is simple: As a general theory for non-Gaussian random fields is currently lacking, Gaussian random fields are the only available choice for an analytical treatment of the current closure problem without involving further phenomenological assumptions. 

Furthermore, Gaussian fields serve as an important reference point for comparison with statistics from real turbulent flow, as for example discussed by \cite{shtilman93jfm,tsinober98ejm,tsinober09book}. It is also worth pointing out that for pressure statistics, the assumption of Gaussianity of the velocity field has, in fact, been shown to lead to qualitatively correct results by \cite{holzer93pfa}. One might speculate that a possible reason for this is that some essential features of the non-locality might be robust to the details of the particular choice of random fields. The motivation for the current work is the perspective that new insights on the structure of the unclosed terms, especially on the complex pressure Hessian, can be achieved this way. In fact, our results will show that nontrivial, but imperfect results can be obtained under the assumption of Gaussian velocity fields.

\subsection{Gaussian characteristic functional for incompressible velocity fields}

The following calculations rely exclusively on the spatial properties of Gaussian random fields, such that we can suppress the time variable in our notation. The assumption of Gaussian velocity fields can most comprehensively be captured on the level of the characteristic functional of the velocity field, which is defined in terms of
\begin{equation}\label{eq:velocitycharfun}
  \phi^u[\bs \lambda(\bs x)] = \left\langle \exp\left[ \mathrm{i}\int\!\mathrm{d}\bs x\, \lambda_i(\bs x) \, u_i(\bs x) \right] \right\rangle \, .
\end{equation}
Here $\bs \lambda(\bs x)$ denotes the Fourier field conjugate to the velocity field. The spatial coordinate $\bs x$ may be regarded as a continuous index of the functional Fourier transform. For Gaussian random fields with zero mean, the explicit expression for the characteristic functional reads
\begin{equation}\label{eq:gaussianvelocitycharfun}
  \phi^u[\bs \lambda(\bs x)] = \exp\left[ -\frac{1}{2}\int\!\mathrm{d}\bs x \int\!\mathrm{d}\bs x' \, \lambda_i(\bs x) R_{ij}^u(\bs x,\bs x') \lambda_j(\bs x') \right] 
\end{equation}
which depends on the velocity covariance tensor $R_{ij}^u(\bs x,\bs x') = \langle u_i(\bs x) \, u_j(\bs x') \rangle$. For homogeneous isotropic turbulence in incompressible flows this tensor takes the form
\begin{equation}
  R_{ij}^u(\bs x,\bs x') = R^u_{ij}(\bs r) = \frac{\langle \bs u^2 \rangle}{3} \left[  f_u(r) \, \delta_{ij} + \frac{1}{2} r f_u'(r) \left[ \delta_{ij} - \hat r_i \hat r_j \right] \right] \, .
\end{equation}
Here, $f_u$  denotes the longitudinal velocity autocorrelation function,  and $\hat r_i=r_i/r$ denotes a component of the direction of the difference vector $\bs r=\bs x-\bs x'$.  We have used the fact that for homogenous turbulence the covariance tensor is a function of the distance vector only, $R^u_{ij}(\bs x,\bs x')=R^u_{ij}(\bs r)$. Both types of notation will be used in the following, depending on the context. The statistics of the Gaussian velocity field is completely specified by the longitudinal velocity autocorrelation function $f_u(r)$. 

Next, we consider the characteristic functional of the velocity gradient tensor, which in analogy to \eqref{eq:velocitycharfun} is defined as
\begin{equation}\label{eq:velgradcharfun}
  \phi^{\mat A}[\mat \Lambda(\bs x)] = \left\langle \exp\left[ \mathrm{i}\int\!\mathrm{d}\bs x \, \Lambda_{ij}(\bs x)A_{ij}(\bs x) \right] \right\rangle \, .
\end{equation}
Owing to $A_{ij}={\partial u_i}/{\partial x_j}$, a simple relation between \eqref{eq:velocitycharfun} and \eqref{eq:velgradcharfun} can be established by partial integration,
\begin{equation}
  \phi^{\mat A}[\mat \Lambda(\bs x)] = \phi^u \left[\bs \lambda(\bs x) = -\nabla \cdot \mat \Lambda^{\mathrm{T}}(\bs x)\right] \, ,
\end{equation}
where the argument of the characteristic functional for the velocity field in component notation reads $\lambda_i(\bs x) = -\frac{\partial \Lambda_{ij}}{\partial x_j}(\bs x)$. This shows that the characteristic functional for the velocity gradient is readily expressed in terms of the characteristic functional of the velocity field. Using this result together with the Gaussian characteristic functional \eqref{eq:gaussianvelocitycharfun} leads, after partial integration, to
\begin{equation}\label{eq:gaussianvelocitygradientcharfun}
  \phi^{\mat A}[\mat \Lambda(\bs x)] = \exp\left[ -\frac{1}{2}\int\!\mathrm{d}\bs x \, \int\!\mathrm{d}\bs x' \, \Lambda_{ik}(\bs x) R_{ijkl}(\bs x,\bs x') \Lambda_{jl}(\bs x') \right] \, .
\end{equation}
Here we have introduced the velocity gradient covariance tensor, which is kinematically related to the velocity covariance tensor:
\begin{equation}\label{eq:velgradcovariance}
  R_{ijkl}(\bs x,\bs x') = \left\langle A_{ik}(\bs x) A_{jl}(\bs x') \right\rangle= \left\langle \frac{\partial u_i}{\partial x_k}(\bs x) \frac{\partial u_j}{\partial x'_l}(\bs x') \right\rangle = \frac{\partial^2 R_{ij}^u}{\partial x_k \partial x'_l}(\bs x,\bs x') \, .
\end{equation}
The general structure of this tensor is discussed in appendix \ref{app:velgradcov}. Two conclusions can be drawn from this result. First, it shows that if the velocity field is (multipoint) Gaussian, so is the velocity gradient tensor field (as cautioned already before, this is known to be unphysical due to intermittency, etc.). 
Second, because the velocity gradient covariance tensor is kinematically prescribed by the velocity covariance tensor, the full statistical description of the velocity gradient tensor field is fixed by one scalar function, the longitudinal velocity autocorrelation function. It is crucial to note that the Reynolds-number dependence as well as any length-scale information of the Gaussian field enter through the correlation function. Its precise shape, for example, especially near the origin, depends on the Reynolds number.

For the calculations of the unclosed terms we will in particular make use of single- and two-point statistics. The characteristic functional contains the statistics for arbitrary numbers of points, such that it can be conveniently projected to the single-point statistics by evaluating \eqref{eq:gaussianvelocitygradientcharfun} at $\mat \Lambda(\bs x)=\mat \Lambda_1 \delta(\bs x - \bs x_1)$:
\begin{equation}\label{eq:singlepointcharfun}
  \phi_1^{\mat A}(\mat \Lambda_1) = \phi^{\mat A}[\mat \Lambda_1 \, \delta(\bs x-\bs x_1)] = \exp\bigg[ -\frac{1}{2}   \mat \Lambda^{\mathrm{T}}_1 \mat R(\bs 0) \mat \Lambda_1 \bigg] \, .
\end{equation}
Here, we have introduced the short-hand notation $\mat \Lambda^{\mathrm{T}}_1 \mat R(\bs 0) \mat \Lambda_1=\Lambda_{1,ik}R_{ijkl}(\bs x_1,\bs x_1)\Lambda_{1,jl}$. The characteristic function for two points is obtained analogously:
\begin{align}\label{eq:twopointcharfun}
  &\phi_2^{\mat A}(\mat \Lambda_1,\mat \Lambda_2) = \phi^{\mat A}[\mat \Lambda_1 \, \delta(\bs x-\bs x_1) + \mat \Lambda_2 \, \delta(\bs x-\bs x_2)] \nonumber \\
  &= \exp\bigg[ -\frac{1}{2}  \big[ \mat \Lambda^{\mathrm{T}}_1 \mat R(\bs 0) \mat \Lambda_1 + \mat \Lambda^{\mathrm{T}}_1 \mat R(\bs r) \mat \Lambda_2 + \mat \Lambda^{\mathrm{T}}_2 \mat R(\bs r) \mat \Lambda_1 + \mat \Lambda^{\mathrm{T}}_2 \mat R(\bs 0) \mat \Lambda_2 \big] \bigg] \, .
\end{align}
It may be noted that both characteristic functions resemble a standard Gaussian characteristic function generalized to tensorial random variables. The velocity gradient tensor PDFs, whenever needed, can be obtained by Fourier transform. Incompressibility is explicitly taken into account by the structure of the covariance tensor (see appendix \ref{app:velgradcov}).

\subsection{Gaussian closure}

\subsubsection{Conditional Laplacian}

We are now equipped with the technical prerequisites to evaluate the unclosed terms based on the assumption of Gaussian random fields. To calculate the conditional viscous term, we first obtain an expression for $\big\langle \mat A(\bs x_2) \big| \A_1 \big\rangle$ and then evaluate the Laplacian by relation \eqref{eq:conditionallaplaciantwopoint}. The details of the calculation are elaborated in appendix \ref{app:firstmomentandlaplacian}, which yields the simple result
\begin{equation}\label{eq:conditionallaplacianmain}
  \big\langle \nu \Delta_{\bs x_1} \mat A(\bs x_1) \big | \A_1 \big\rangle = \delta \A_1
\end{equation}
with
\begin{equation}
  \delta = \nu \frac{7}{3} \, \frac{f_u^{(4)}(0)}{f_u''(0)} = -\nu  \frac{\int \! \mathrm{d}k \, k^4 \, E(k)}{\int \! \mathrm{d}k \, k^2 \, E(k)}  \, , \label{eq:deltamaintext}
\end{equation}
where $E(k)$ is the energy spectrum function. That means, for incompressible Gaussian velocity fields, the conditional Laplacian is a linear function of the velocity gradient with a prefactor depending on the kinematic viscosity as well as on derivatives of the longitudinal autocorrelation function of the velocity, or alternatively on integrals involving the energy spectrum function. It is interesting to note that the Gaussian closure is consistent with the linear diffusion model by \cite{martin98pof}, thus giving theoretical underpinning for this phenomenological assumption: For the case of Gaussian velocity fields, the assumption of a linear dependence of the conditional Laplacian is exact. In addition to the linear diffusion model, however, the Gaussian closure also fixes the coefficient by expressing it in terms of the longitudinal velocity autocorrelation function or, equivalently, the energy spectrum function.

The conditional Laplacian depends implicitly on the Reynolds number through the autocorrelation function. As a simple estimate of the Reynolds-number dependence of this term one may note that the coefficient $\delta$ defines the inverse of a time scale. If we assume, in accordance with Kolmogorov's phenomenology, that this time scale depends on the mean rate of energy dissipation $\varepsilon$ and viscosity $\nu$, it is necessarily identified with the Kolmogorov time scale $\tau_{\eta}=(\nu/\varepsilon)^{1/2}$. The same can be concluded from \eqref{eq:deltamaintext} since both integrals in the numerator and denominator are dominated by viscous scales. An order-of-magnitude estimate based on a simple model spectrum described in appendix \ref{app:modelspectrum} yields $\delta \, \tau_{\eta}\approx -0.65$ for a range of Reynolds numbers. This value will turn out to be excessive in magnitude because of an imperfectly modeled viscous range. We will discuss a more accurate estimate based on the velocity derivative skewness in the context of the enhanced Gaussian closure later below.

\subsubsection{Conditional pressure Hessian}

To evaluate the non-local contribution to the pressure Hessian under the assumption of incompressible Gaussian velocity fields, we first obtain the Gaussian expression for $\big\langle Q(\bs x_2) \big | \A_1 \big \rangle$ and insert this into \eqref{eq:conditionalpressurehessianmain}. The details of this calculation are included in appendix \ref{app:secondmomentandhessian}, so that we can here focus on a discussion of the main result. It takes the form
\begin{align}
  \big\langle \mat{\widetilde{H}}(\bs x_1) \big| \A_1 \big\rangle &=  \alpha \, \left( \S_1^2 - \frac{1}{3}\mathrm{Tr}\left( \S_1^2 \right) \mat I \right) \nonumber \\
                      									    &+  \beta \, \left( \W_1^2 - \frac{1}{3}\mathrm{Tr}\left( \W_1^2 \right) \mat I \right) \nonumber \\
									       		    &+  \gamma \, \left( \S_1\W_1 - \W_1\S_1  \right) \label{eq:nonlocalpressurehessianmaintext}
\end{align}
with
\begin{subequations}\begin{align}
  \alpha &= -\frac{2}{7} \\
  \beta &= -\frac{2}{5} \\
  \gamma &=  \frac{6}{25} + \frac{16}{75 f_u''(0)^2} \int \! \mathrm{d}r \, \frac{f_u'f_u'''}{r} \, . \label{eq:gammamaintext}
\end{align}\end{subequations}

First, it is interesting to note that the non-local contribution to the conditional pressure Hessian consists of all possible combinations of the rate-of-strain tensor $\S$ and the rate-of-rotation tensor $\W$ that are dimensionally consistent (quadratic), traceless and symmetric. Second, the fact that the local nonlinear term in the stochastic evolution equation \eqref{eq:stochasticcharacteristic} in terms of the rate-of-strain and the rate-of-rotation tensors takes the form
\begin{equation}
  -\left( \A_1^2 - \frac{1}{3} \mathrm{Tr}\left( \A_1^2 \right) \mat I \right) = -\left( \S_1^2 - \frac{1}{3} \mathrm{Tr}\left( \S_1^2 \right) \mat I \right)-\left( \W_1^2 - \frac{1}{3} \mathrm{Tr}\left( \W_1^2 \right) \mat I \right)- \S_1\W_1-\W_1\S_1
\end{equation}
shows that the effect of the Gaussian non-local pressure Hessian can, in part, be understood in terms of a ``reduction of nonlinearity" as phenomenologically introduced by \cite{chertkov99pof} and \cite{naso05pre} for the perceived velocity gradient tensor dynamics based on Lagrangian tetrads. This  result from the Gaussian approximation gives further theoretical support for such an Ansatz. The Gaussian approximation, however, yields differing coefficients for the rate-of-strain term and the rate-of-rotation term, thus suggesting a refinement of the reduction of nonlinearity. 

A remarkable property of the restricted Euler approximation is that the velocity gradient invariants dynamics can be reduced to only two invariants, $R$ and $Q$, out of five possible invariants \citep{vieillefosse82jpf,cantwell92pfa}. This is a direct consequence of the fact that only the isotropic, local contribution of the pressure Hessian (which involves the trace of the squared velocity gradient) is taken into account in this approximation. In comparison to that, the Gaussian approximation also incorporates deviatoric contributions, which cannot be expressed in terms of the identity tensor or powers of $\A$ only. This results in a dynamical system that cannot be reduced to $R$ and $Q$ only. In that sense the structure of the non-local pressure Hessian in the Gaussian approximation suggests the possibility of more complex dynamics compared to the restricted Euler model.

Regarding the Reynolds-number dependence, it is interesting to note that the coefficients $\alpha$ and $\beta$ do not depend on the autocorrelation function and thus are independent of the Reynolds number.  The situation is more involved for the coefficient $\gamma$ due to its integral dependence on the velocity autocorrelation function. A closer look at this integral relation, however, reveals that it depends on the non-dimensional function $f_u'(r)f_u'''(r)/f_u''(0)^2$. Composed of derivatives of the velocity autocorrelation function, it is plausible to assume that this function is largely independent of the energy-containing range of turbulent motion. If we additionally assume that the inertial and dissipative ranges of the energy spectrum exhibit a universal functional form at sufficiently high Reynolds numbers, this implies a universal shape of this non-dimensional function when properly rescaled by the  Kolmogorov length scale $\eta$. Under these assumptions the coefficient $\gamma$ is also universal, i.e. Reynolds-number independent. The actual numeric value of $\gamma$, however, must be obtained from a model autocorrelation function. We have used the model spectrum described in appendix \ref{app:modelspectrum} to estimate the parameter $\gamma$ for a Reynolds number of $R_{\lambda}\approx432$ and obtained $\gamma\approx 0.08$. Also the universality argument was verified for a range of Reynolds numbers. Reynolds-number-dependent effects like intermittency and bottleneck corrections as discussed, e.g., by \cite{meyers08pof}, however, will break the universality of the autocorrelation function and the coefficient $\gamma$.

\section{Dynamics of the Gaussian closure}

\subsection{General considerations}
With the results of the last section we have obtained a closure to the statistical evolution equation \eqref{eq:stochasticcharacteristic}, such that we arrive at a stochastic differential equation (SDE),
\begin{align} \label{eq:closurefullmodel}
  \mathrm{d}\A = \bigg[&-\left(\A^2-\frac{1}{3}\mathrm{Tr}\left( \A^2 \right) \mat I \right) -\alpha \, \left( \S^2 - \frac{1}{3}\mathrm{Tr}\left( \S^2 \right)  \mat I \right) \nonumber \\
                        &-\beta \, \left( \W^2 - \frac{1}{3}\mathrm{Tr}\left( \W^2 \right) \mat I \right) - \gamma \, \left( \S\W - \W\S  \right)  + \delta \A \bigg] \mathrm{d}t + \mathrm{d}\mat F \, ,
\end{align}
where the coefficients $\alpha$ to $\delta$ are fixed by the Gaussian approximation. We have dropped the subscripts as all quantities are considered at a single point. This Langevin equation describes the evolution of a class of fluid particles which share the same value of the velocity gradient.

For the interpretation of the individual terms it turns out to be useful to decompose the evolution equation for $\A$ into the dynamics of the rate-of-strain tensor $\S$ and rate-of-rotation tensor $\W$:
\begin{align}
  \mathrm{d}\S = \bigg[ -(1+\alpha) \, \left( \S^2 - \frac{1}{3}\mathrm{Tr}\left( \S^2 \right)  \mat I \right) &-(1+\beta) \, \left( \W^2 - \frac{1}{3}\mathrm{Tr}\left( \W^2 \right) \mat I \right) \nonumber \\ 
  &- \gamma \, \left( \S\W - \W\S  \right)  + \delta \S \bigg] \mathrm{d}t + \mathrm{d}\mat F_{\mat S}
\end{align}
and
\begin{equation}
  \mathrm{d}\W = \left[ -\S\W-\W\S + \delta \W \right] \mathrm{d}t + \mathrm{d}\mat F_{\mat W} \, .
\end{equation}
Here, $\mathrm{d}\mat F_{\mat S}$ and $\mathrm{d}\mat F_{\mat W}$ denote the contributions of the stochastic forcing to the rate-of-strain and rate-of-rotation dynamics. The rate-of-rotation dynamics written in terms of the vorticity takes the form
\begin{equation}
  \mathrm{d}\bs \omega = \left[ \mat S \bs \omega + \delta \bs \omega \right] \mathrm{d}t + \mathrm{d}\bs F_{\bs \omega} \, .
\end{equation}

To elucidate the dynamics of the closure, we take the restricted Euler model as a reference, whose dynamical properties have been discussed in detail by \cite{cantwell92pfa} and \cite{nomura98jfm}. For the vorticity equation, the well-known effect of vortex stretching is supplemented by a linear damping term and the random-force term.

The same applies to the rate-of-strain equation. Additionally, the terms related to $\alpha$ and $\beta$ modify the coefficients of the restricted Euler model, i.e. their relative strength and correspondingly their relative time scale is changed. As we find typical values in the range $-1 < \alpha < 0$ and $-1 < \beta < 0$, the effect of the restricted Euler terms is weakened. It is plausible to assume that this affects the occurrence of a finite-time singularity.

It is furthermore instructive to consider the effect of the terms individually. The $\alpha$-term in the rate-of-strain equation can be diagonalized along with the rate-of-strain tensor and thus affects only the eigenvalues of $\S$ and not the orientation of the eigenframe. It may be understood as an algebraic growth or decay of the eigenvalues, which additionally maintains the zero-trace condition of the rate-of-strain tensor. Considered on its own, it leads to a divergence in finite time, similar to that observed in the inviscid Burgers equation and the restricted Euler model.

The $\beta$-term is independent of the rate-of-strain tensor, and thus depends linearly on time for a fixed vorticity. Considered in a frame in which the $\bs e_1$-axis coincides with the direction of the vorticity, this term is diagonal and induces a linear decrease of the first diagonal component of $\S$ at a rate of $-(\beta+1)\bs \omega^2/6$ and an increase of the remaining two diagonal elements at a rate of $(\beta+1) \bs \omega^2/12$, assuming $\beta>-1$. As an accompanying effect, the eigendirection corresponding to the most negative eigenvalue tends to align with the direction of the vorticity.
It turns out that this term is crucial for preventing a blow-up of the SDE system. As a dynamically evolving vorticity tends to be amplified along the direction of the most positive eigenvalue of $\S$, the $\beta$-term counteracts an unbounded growth because it decreases the most positive eigenvalue with a rate $\sim \bs \omega^2$, even allowing it to become negative eventually. This goes along with a tilting of the eigenframe of $\S$.

The $\gamma$-term is a new term not present in the restricted Euler model. Considered for a fixed vorticity, it leads to a rotation of the eigenframe of $\S$ without changing its eigenvalues. This can be seen by the fact that a rotation of $\S$ with a rotation matrix $\mat M$ takes the form $\S' = \mat M \S \mat M^{\mathrm{T}}$. For an infinitesimal time interval the rate-of-rotation tensor induces a rotation of the form $\mat M = \mat I + \W \mathrm{d}t$, such that the infinitesimal evolution of $\S$ takes the form
\begin{equation}
  \mathrm{d}\S = -\left[ \S\W - \W\S \right] \mathrm{d}t
\end{equation}
which is precisely the term associated with $\gamma$. Combined with a vorticity vector that undergoes vortex stretching with the tendency to align with the direction of the most positive eigenvector, the $\gamma$-term induces an accelerated rotation of the eigenframe of $\S$ about this axis. The full system \eqref{eq:closurefullmodel} shows a complex dynamics, whose statistics will be discussed below.

Finally, we study the magnitudes of the different terms and their Reynolds-number dependence. To this end it is useful to non-dimensionalize \eqref{eq:closurefullmodel}. The velocity gradient tensor being a small-scale quantity motivates a non-dimensionalization with the Kolmogorov time scale $\tau_{\eta} = \left(2\langle \mathrm{Tr}\left( \mat S^2 \right) \rangle \right)^{-1/2}$. With $\A_{\star}=\A\tau_{\eta}$ and $t_{\star}=t/{\tau_{\eta}}$, \eqref{eq:closurefullmodel} takes the form
\begin{align} \label{eq:closurefullmodelnondimensionalized}
  \mathrm{d}\A_{\star} = \bigg[&-\left(\A_{\star}^2-\frac{1}{3}\mathrm{Tr}\left( \A_{\star}^2 \right) \mat I \right) -\alpha \, \left( \S_{\star}^2 - \frac{1}{3}\mathrm{Tr}\left( \S_{\star}^2 \right)  \mat I \right) \nonumber \\
                        &-\beta \, \left( \W_{\star}^2 - \frac{1}{3}\mathrm{Tr}\left( \W_{\star}^2 \right) \mat I \right) - \gamma \, \left( \S_{\star}\W_{\star} - \W_{\star}\S_{\star}  \right)  + \delta_{\star} \A_{\star} \bigg] \mathrm{d}t_{\star} + \mathrm{d}\mat F_{\star} \, ,
\end{align}
where $\delta_{\star}=\delta\tau_{\eta}$ and $\mathrm{d}\mat F_{\star} = \mathrm{d}\mat F \tau_{\eta}$. The non-dimensional coefficients $\alpha$, $\beta$ and $\gamma$ remain unchanged by this procedure. As discussed above, $\alpha$ and $\beta$ are Reynolds-number independent and the same is expected for $\gamma$ under the simple universality argument. 

As will be derived from the Kolmogorov equation in section \ref{sec:dnsestimates}, we have $\delta_{\star}=7S_3 / (6 \sqrt{15}) $, where $S_3$ is the derivative skewness coefficient (typically $S_3\approx-0.5$ with  a weak dependence on Reynolds number \citep{frisch95book} due to small-scale intermittency which we do not take into account in this work). Thus also the linear damping term may vary weakly with the Reynolds number. As a consequence, the deterministic part of the velocity gradient tensor dynamics exhibits only a minor variation with the Reynolds number in the realm of the Gaussian approximation when non-dimensionalized with the Kolmogorov time scale. This is different, however, for the large-scale forcing term: If we keep the large-scale time scale imposed by the forcing term fixed, the Kolmogorov time scale decreases with increasing Reynolds number. As a result the non-dimensional forcing term becomes smaller with increasing Reynolds number. This appears plausible because the small-scale statistics should become independent of the non-universal large-scale forcing for sufficiently high Reynolds numbers.

\subsection{Shortcomings of the Gaussian closure}

\begin{figure}
\begin{center}
  \includegraphics[width=1.0\textwidth]{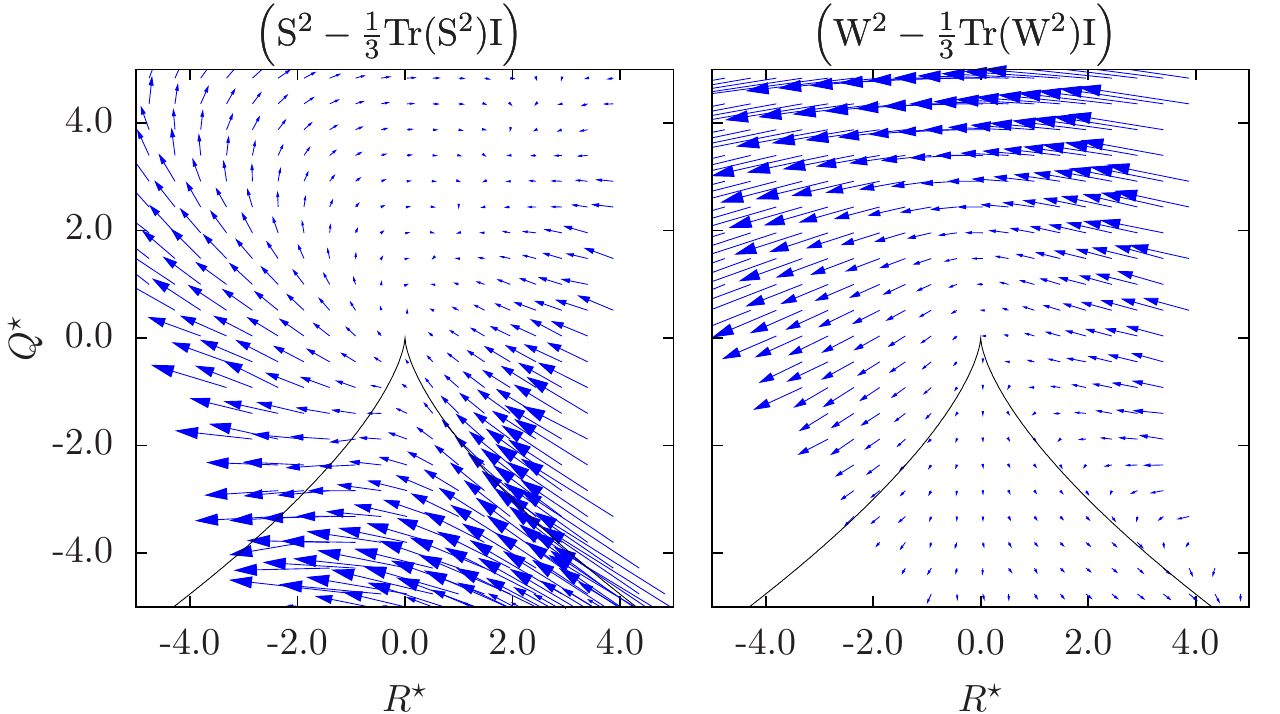}
\end{center}
\caption{$\left(\mat S^2 - \frac{1}{3}\mathrm{Tr}\left( \mat S^2 \right) \mat I \right)$ and $\left( \mat W^2 - \frac{1}{3}\mathrm{Tr}\left( \mat W^2 \right) \mat I \right)$ projected to the $RQ$-plane (see text for details on the projection). The strain term is predominantly active in the strain-dominated region $Q^{\star}<0$ whereas rotation term is predominantly active in the vorticity-dominated region $Q^{\star}>0$. All conditional averages have been non-dimensionalized by $\big\langle \mathrm{Tr}\left( \mat S^2 \right) \big\rangle$.  Here, and for the following plots, vectors have been scaled with a factor of $0.08$.}
\label{fig:rq_vecfield_s2_w2}
\end{figure}

\begin{figure}
\begin{center}
  \includegraphics[width=1.0\textwidth]{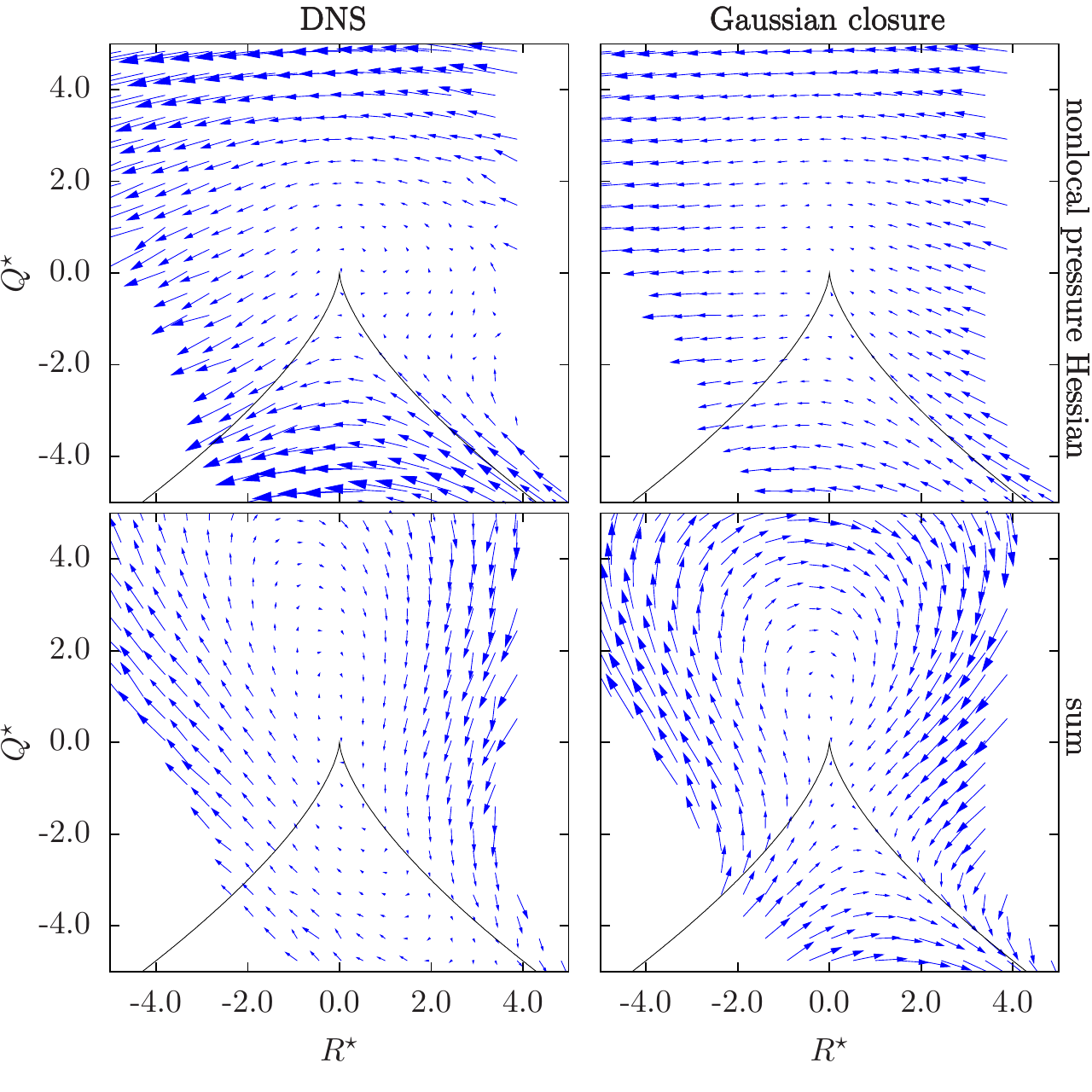}
\end{center}
\caption{Top panels: Non-local pressure Hessian contributions from DNS and the Gaussian closure, projected to the $RQ$-plane. Bottom panels: Sum of the restricted Euler, non-local pressure Hessian and viscous contributions from DNS and the Gaussian closure.} \label{fig:rq_vecfield_panels2}
\end{figure}

The evolution of the full system \eqref{eq:closurefullmodel}, in which all of the terms are simultaneously active and completed by the (stochastic) forcing term, is governed by a combination of all effects discussed above, which can lead to complex temporal behaviour. It is plausible  that whether or not the system diverges depends crucially on the particular choice of the parameters $\alpha$ to $\delta$. If we assume that the singularity of the restricted Euler model should be regularized by the non-local pressure Hessian contributions, $\alpha$, $\beta$ and $\gamma$ are the crucial parameters. As we have seen, $\alpha$ and $\beta$ are universal, whereas $\gamma$ depends on the two-point correlation of the velocity field.  

To further elucidate the question of stability of the closure, let us consider first the projections of the different terms to the $RQ$-plane. Here and in the following, we use velocity gradient tensor fields from DNS to evaluate the terms of the closure, contract them with $-\mat A$ and $-\mat A^2$, respectively, and non-dimensionalize them with proper powers of $\left\langle \mathrm{Tr}\left (\mat S^2 \right) \right\rangle$. Finally, they are binned with respect to $Q^{\star} = Q/\left\langle \mathrm{Tr}\left( \mat S^2 \right) \right\rangle$ and $R^{\star} = R/\left\langle \mathrm{Tr}\left( \mat S^2 \right) \right\rangle^{3/2}$. With this procedure, the contribution of the different terms to the reduced $RQ$-dynamics is revealed, which helps to compare our closure, for example, to the restricted Euler model. It has to be stressed, though, that the dynamics of the closure involves all five possible invariants, i.e. information is lost due to projection. In this context, we would like to refer the reader to the recent work by \cite{luehti09jfm} for a study of the velocity gradient dynamics beyond the $RQ$-plane.

We analyse data from the JHU turbulence database \citep{li08jot}, where data from a DNS at a Taylor-scale Reynolds number of $R_{\lambda}\approx 433$ is publicly available. For the following estimates a single $1024^3$ snapshot of the database was used. However, we also performed further checks on a number of snapshots to ensure that the reported results are robust. Velocity gradients and the Laplacian of the velocity gradients have been calculated spectrally from the downloaded velocity fields. Due to limited resolution of the smallest scales, the velocity field was low-pass filtered for the evaluation of the velocity gradients and the Laplacian of the velocity gradients. The cut-off was determined such that the unphysical pile-up of kinetic energy at the highest wavenumbers (i.e. beyond approximately $0.85 \, k_{\mathrm{max}}$, where $k_{\mathrm{max}}$ denotes the highest dynamically active wavenumber) is excluded. As a consequence we expect inaccuracies especially for the viscous contributions. The pressure Hessian has been evaluated spectrally from the downloaded pressure fields.

To discriminate the effects of the terms related to $\alpha$ and $\beta$, figure \ref{fig:rq_vecfield_s2_w2} shows the projections of $\left( \mat S^2 - \frac{1}{3}\mathrm{Tr}\left( \mat S^2 \right) \mat I \right)$ and $\left( \mat W^2 - \frac{1}{3}\mathrm{Tr}\left( \mat W^2 \right) \mat I \right)$ to the $RQ$-plane, respectively. The $\alpha$-term turns out to be predominantly active in the strain-dominated half-plane ($Q^{\star}<0$) whereas the converse is true for the $\beta$-term. Interestingly, the term associated with $\gamma$ does not contribute to this projection, because $\mat S \mat W - \mat W \mat S$ vanishes when contracted with arbitrary powers of $\mat A$.

The top panel of figure \ref{fig:rq_vecfield_panels2} compares the non-local pressure Hessian contribution from DNS with the Gaussian closure. While the overall topology is quite similar, differences in magnitude can be observed. As in the DNS, the closure attenuates the singularity induced along the right part of the Vieillefosse line, however less strongly. This becomes even more clear when comparing the sum of the contributions from the restricted Euler term, the pressure Hessian and the viscous contributions from DNS to the Gaussian closure, as shown in the bottom panel of figure \ref{fig:rq_vecfield_panels2}. For the Gaussian closure, the total effect is apparently not strong enough to counteract the singularity induced along the right Vieillefosse tail. Because projection to the $RQ$-plane does not allow to comprehensively judge the performance of this closure, we also ran numerical tests solving \eqref{eq:closurefullmodelnondimensionalized} in the unforced, inviscid case with the parameter values given by the Gaussian closure. This consistently led to blow-up, i.e. the non-local pressure Hessian contributions in the Gaussian closure are not able to prevent  singularities by themselves. Also including the linear damping term for the parameter range discussed below does not solve this problem. Tests show that the system still diverges for certain initial conditions, which is consistent with the \textit{a priori} findings reported in figure \ref{fig:rq_vecfield_panels2}. This is because the linear damping term becomes subdominant for sufficiently large velocity gradients compared to the quadratic terms of the restricted Euler and non-local pressure contributions \citep{meneveau11arf}.

We conclude that, while the structure of the Gaussian closure yields promising insights, it fails to predict coefficients that lead to a non-divergent evolution in time.  A further numerical analysis of how the system diverges shows that especially the reduction of nonlinearity associated with the $\alpha$-term is not sufficient to prevent the singularity, which can be regarded as the main shortcoming of the Gaussian closure. The spatial structure of Gaussian random fields is insufficient to counteract the local self-amplification effects by the restricted Euler part of the dynamics.

\subsection{Estimation of the coefficients from DNS data}
\label{sec:dnsestimates}

To overcome the shortcomings of the Gaussian approximation, we accept the tensorial structure of the non-local pressure Hessian contributions as well as the conditional Laplacian as obtained analytically from the Gaussian closure, but obtain estimates for the parameter values from DNS data. We will refer to this as the enhanced Gaussian closure. 

To estimate the parameters from the DNS data, where the conditional averages of tensor elements can be directly measured, we recognize that 
in general we have an overdetermined system if we wish to obtain scalar parameters. The most practical direct manner is to consider tensor contractions with appropriate powers of the velocity gradient tensor such that non-vanishing moments can be constructed.

For the linear Laplacian term one may consider the expectation value of the contraction of  the velocity gradient with its Laplacian, 
\begin{equation}\label{eq:contractionlaplacian}
  \int\!\mathrm{d}\A \,\big\langle \nu \mathrm{Tr}\left( \mat A^{\mathrm{T}}\Delta \mat A\right) \big | \A \big\rangle \, f(\A) =   \delta \int\!\mathrm{d}\A \, \mathrm{Tr}\left( \A^{\mathrm{T}}\A \right) \, f(\A)
\end{equation}
which yields the relation
\begin{equation}
  \delta = \frac{\big\langle \nu \mathrm{Tr}\left( \mat A^{\mathrm{T}}\Delta \mat A\right) \big\rangle}{\big\langle \mathrm{Tr}\left( \mat A^{\mathrm{T}}\mat A\right) \big\rangle} \, .
\end{equation}
It can be shown by a straightforward calculation from the velocity gradient covariance structure \eqref{eq:covariancestructure} that this estimate of $\delta$ is equivalent to the result \eqref{eq:deltamaintext} from the Gaussian approximation. Evaluation of the averages in the numerator and denominator from the DNS data leads to the result $\delta \tau_{\eta} \approx -0.15$.

To provide further theoretical insight, $\delta$ can be related to the velocity derivative skewness. The  relevant derivation starts from the Kolmogorov equation \citep{monin07book2} for homogeneous isotropic turbulence
\begin{equation}
  \left\langle \left(u_1({\bs x} + r{\bs e}_1) - u_1({\bs x})\right)^3 \right\rangle - 6 \nu \frac{\mathrm{d}}{\mathrm{d}r}  \left\langle \left(u_1({\bs x} + r{\bs e}_1) - u_1({\bs x})\right)^2 \right\rangle = - \frac{4}{5} \varepsilon r \, .
\end{equation}
An expansion of this expression about $r=0$ yields
\begin{equation}
\left\langle\left(\frac{\partial u_1}{\partial x_1}\right)^3\right\rangle r^3 - 12 \nu  \left\langle\left(\frac{\partial u_1}{\partial x_1} \right)^2\right\rangle  r + 2 \nu   \left\langle \left(\frac{\partial^2  u_1}{\partial x_1^2} \right)^2 \right\rangle ~ r^3 + \mathrm{h.o.t.} = - \frac{4}{5} \varepsilon r \, .
\end{equation}
Comparing the terms of order one yields the usual relation $\varepsilon = 15 \nu  \left\langle\left( {\partial u_1}/{\partial x_1} \right)^2\right\rangle$. The terms of third order give
\begin{equation}
  \left\langle \left(\frac{\partial u_1}{\partial x_1}\right)^3\right\rangle = -2 \nu \left\langle \left( \frac{\partial^2  u_1}{\partial x_1^2} \right)^2 \right\rangle \, ,
\end{equation}
which can be used to relate the fourth-order derivative in \eqref{eq:deltamaintext} to the velocity gradient skewness $S_3=\left\langle(\partial u_1/\partial x_1)^3\right\rangle/\left\langle(\partial u_1 / \partial x_1)^2\right\rangle^{3/2}$. With $\left\langle \left({\partial u_1}/{\partial x_1}\right)^2\right\rangle = -(\langle \bs u^2\rangle/3) f_u''(0)$ and $\left\langle \left( {\partial^2  u_1}/{\partial x_1^2} \right)^2 \right\rangle = (\langle \bs u^2\rangle/3) f_u^{(4)}(0)$ the coefficient $\delta$ for isotropic turbulence obeying the Kolmogorov equation takes the form
\begin{equation} 
  \delta = \frac{7}{6} {\left\langle \left(\frac{\partial u_1}{\partial x_1}\right)^3 \right\rangle} \bigg / {\left\langle \left(\frac{\partial u_1}{\partial x_1}\right)^2 \right\rangle} = \frac{7}{6 \sqrt{15}} \frac{S_3}{\tau_{\eta}} \, .
\end{equation}
As a result the conditional Laplacian depends on the Reynolds number mainly through $\tau_{\eta}$, but also due to a weak dependence of $S_3$. 
The derivative skewness is known to have values near $S_3 \approx -0.5$. Recent DNS results by \cite{ishihara07jfm} at $R_{\lambda}=471$, for example, provide a value of $S_3$ between $-0.6$ and $-0.5$, which leads to a value of $\delta \tau_{\eta}$ between $-0.15$ and  $-0.18$, which is in good agreement with our DNS findings. For further analysis, the value of  $\delta \tau_{\eta} \approx -0.15$ will be assumed. 

To estimate the coefficients of the non-local pressure Hessian contribution \eqref{eq:nonlocalpressurehessianmaintext} from DNS, we obtain expectation values analogous to \eqref{eq:contractionlaplacian}. The lowest-order contraction that leads to a non-vanishing of the $\gamma$-term is the contraction with the quadratic expression $\mat S \mat W$. For consistency, we also use contractions with  quadratic expressions of the form $\mat S^2$ and $\mat W^2$ to estimate $\alpha$ and $\beta$, which leads to a linear set of equations. This procedure is also interesting from a physical point of view, because it connects the parameter estimates for the non-local pressure Hessian with the flatness factors of the strain field and the vorticity field. The set of equations is readily solved by
\begin{align}
  \alpha &= \frac{  \big\langle\mathrm{Tr}\big( \mat W^2 \widetilde{\mat W^2} \big) \big\rangle \big\langle \mathrm{Tr}\big( \mat S^2 \widetilde{\mat H} \big) \big\rangle - \big\langle \mathrm{Tr}\big( \mat S^2 \widetilde{\mat W^2} \big) \big\rangle \big\langle \mathrm{Tr}\big( \mat W^2 \widetilde{\mat H} \big) \big\rangle }
  { \big\langle \mathrm{Tr}\big( \mat S^2 \widetilde{\mat S^2} \big) \big\rangle \big\langle \mathrm{Tr}\big( \mat W^2 \widetilde{\mat W^2} \big) \big\rangle - \big\langle \mathrm{Tr}\big( \mat S^2 \widetilde{\mat W^2} \big) \big\rangle \big\langle \mathrm{Tr}\big( \mat W^2 \widetilde{\mat S^2} \big) \big\rangle } \\
  \beta &= \frac{   \big\langle\mathrm{Tr}\big( \mat S^2 \widetilde{\mat S^2} \big) \big\rangle \big\langle \mathrm{Tr}\big( \mat W^2 \widetilde{\mat H} \big) \big\rangle - \big\langle \mathrm{Tr}\big( \mat W^2 \widetilde{\mat S^2} \big) \big\rangle \big\langle \mathrm{Tr}\big( \mat S^2 \widetilde{\mat H} \big) \big\rangle }{ \big\langle \mathrm{Tr}\big( \mat S^2 \widetilde{\mat S^2} \big) \big\rangle \big\langle \mathrm{Tr}\big( \mat W^2 \widetilde{\mat W^2} \big) \big\rangle - \big\langle \mathrm{Tr}\big( \mat S^2 \widetilde{\mat W^2} \big) \big\rangle \big\langle \mathrm{Tr}\big( \mat W^2 \widetilde{\mat S^2} \big) \big\rangle } \\
  \gamma &= \frac{ \big\langle \mathrm{Tr}\big( \mat S \mat W \widetilde{\mat H} \big) \big\rangle}{ \big\langle \mathrm{Tr}\big( \mat S \mat W [ \mat S \mat W - \mat W \mat S ] \big) \big\rangle } \, .
\end{align}
We have used the short-hand notation $\widetilde{\mat W^2}=\mat W^2-\frac{1}{3}\mathrm{Tr}\left( \mat W^2 \right)\mat I$ and $\widetilde{\mat S^2}=\mat S^2-\frac{1}{3}\mathrm{Tr}\left( \mat S^2 \right)\mat I$ for denoting the traceless tensors. This means that the parameters $\alpha$, $\beta$ and $\gamma$ can be fixed by evaluating averages involving the rate-of-strain and rate-of-rotation tensors as well as the pressure Hessian.

From the DNS data set we obtain $\alpha\approx-0.61$, $\beta\approx-0.65$, and $\gamma\approx 0.14$, i.e. all values are larger in magnitude than predicted by the Gaussian approximation. In particular, this has the consequence that the nonlinearity associated with the $\alpha$-term is weaker compared to the restricted Euler model and the Gaussian closure. As we will see below, this will allow for a non-divergent evolution of the system. 

Our tests with different snapshots of the DNS data indicate that the reported values are accurate within a few per cent. They may, however, show a Reynolds-number dependence, whose study goes beyond the scope of the current work. We close this section with reiterating that the only difference between the Gaussian closure and its enhanced version is in the choice of the constant coefficients of the non-local pressure Hessian.

\subsection{Comparison of enhanced Gaussian closure and DNS results}
\label{sec:comparison}
\begin{figure}
\begin{center}
  \includegraphics[width=0.85\textwidth]{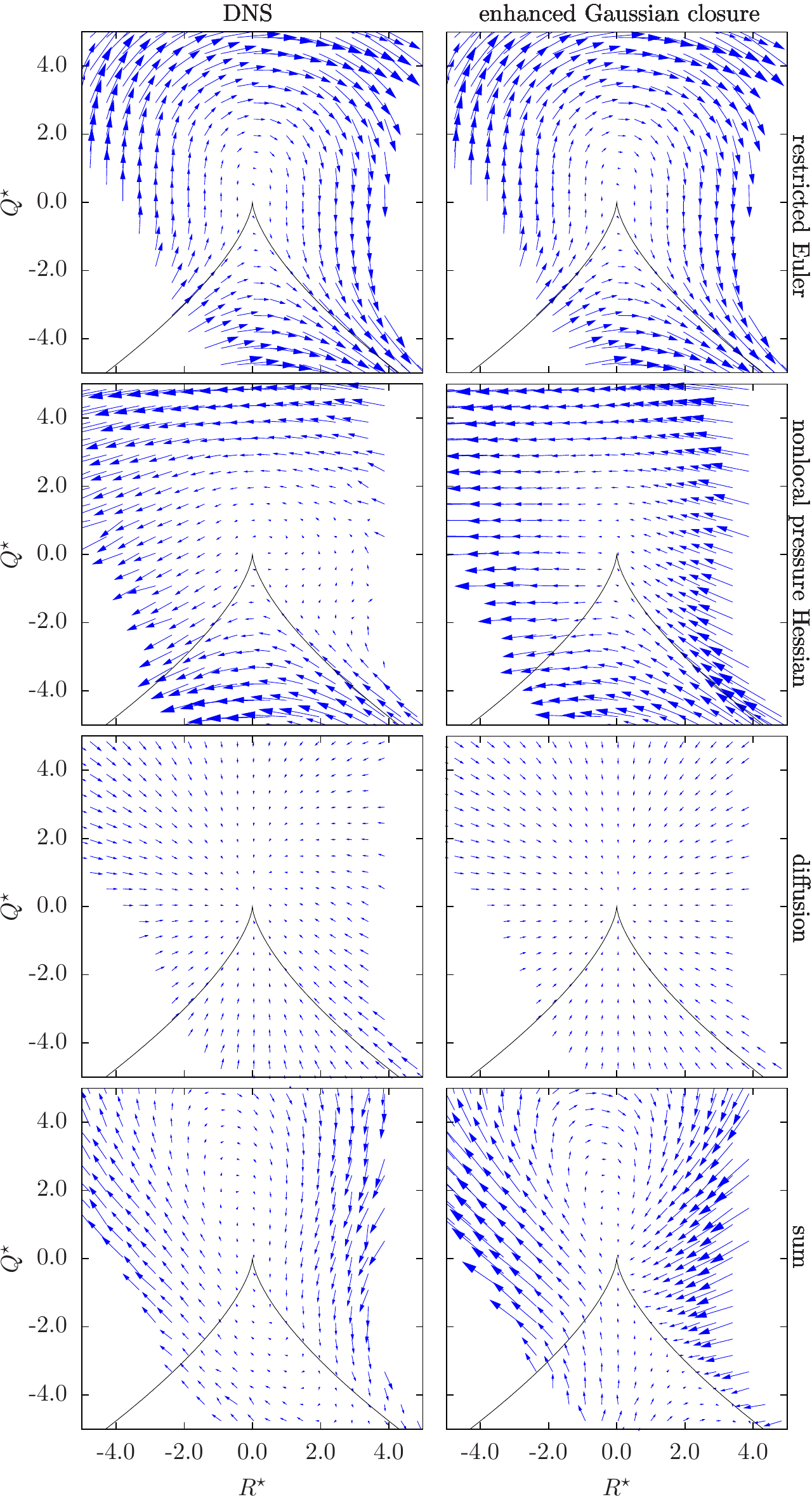}
\end{center}
\caption{Projection of the terms governing the $RQ$-evolution from DNS data and the enhanced Gaussian closure.} \label{fig:rq_vecfield_panels}
\end{figure}

Figure \ref{fig:rq_vecfield_panels} shows a comparison of the different terms of the velocity gradient dynamics estimated from the JHU DNS database along with the terms of the enhanced Gaussian closure closure projected to the $RQ$-plane. As the restricted Euler contribution appears closed in this description, the closure coincides with the DNS data.

The non-local pressure Hessian contribution of the enhanced Gaussian closure shares a number of qualitative similarities with the term from DNS data. Most importantly, and in contrast to the original Gaussian closure, it fully counteracts the singularity along the right branch of the Vieillefosse line. It also reproduces the tendency of the pressure Hessian to push towards negative $R^{\star}$ for positive $Q^{\star}$. The main difference is observed for positive $R^{\star}$ and moderate values of $Q^{\star}$, where the closure does not reproduce the very small amplitudes of the non-local pressure Hessian from DNS.

The diffusive term of the closure compares quite well to the diffusive term from DNS in its overall damping influence, although minor differences in direction and amplitude can be observed. We note that the enhanced Gaussian closure coincides with the original one in this term.

The total vector field of the enhanced closure compares qualitatively well with the DNS results; it indicates a clockwise cyclic motion in the $RQ$-plane. However, the enhanced closure seems to push towards the right Vieillefosse tail too early, such that the PDF of $R^{\star}$ and $Q^{\star}$ will show discrepancies in this region.

\subsection{Enhanced Gaussian closure in SDE}

\begin{figure}
\begin{center}
  \includegraphics[width=0.65\textwidth]{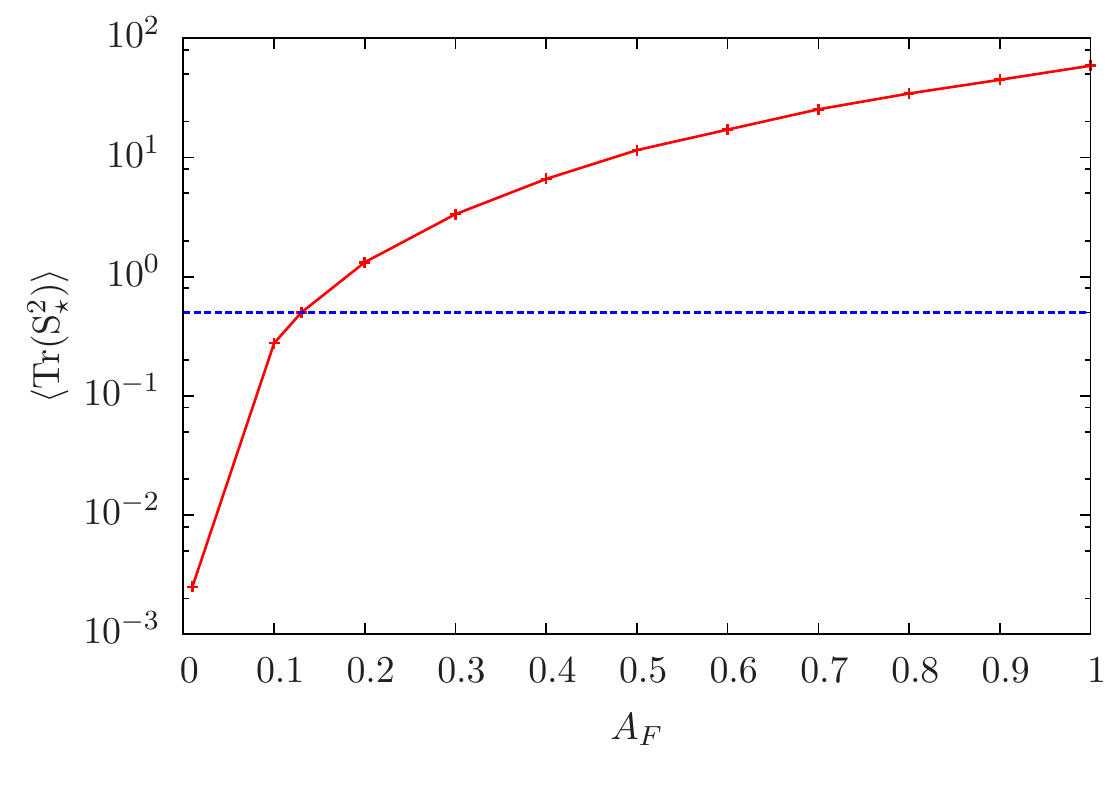}
\end{center}
\caption{$\left\langle \mathrm{Tr}\left( \mat S_{\star}^2 \right) \right\rangle$ as a function of the forcing amplitude $A_F$ for the enhanced Gaussian closure SDE. The dashed blue line indicates the constraint $\left\langle \mathrm{Tr}\left( \mat S_{\star}^2 \right)\right\rangle=1/2$, which is used to determine the amplitude.} \label{fig:force_amplitude}
\end{figure}

\begin{figure}
\begin{center}
   \includegraphics[width=0.49\textwidth]{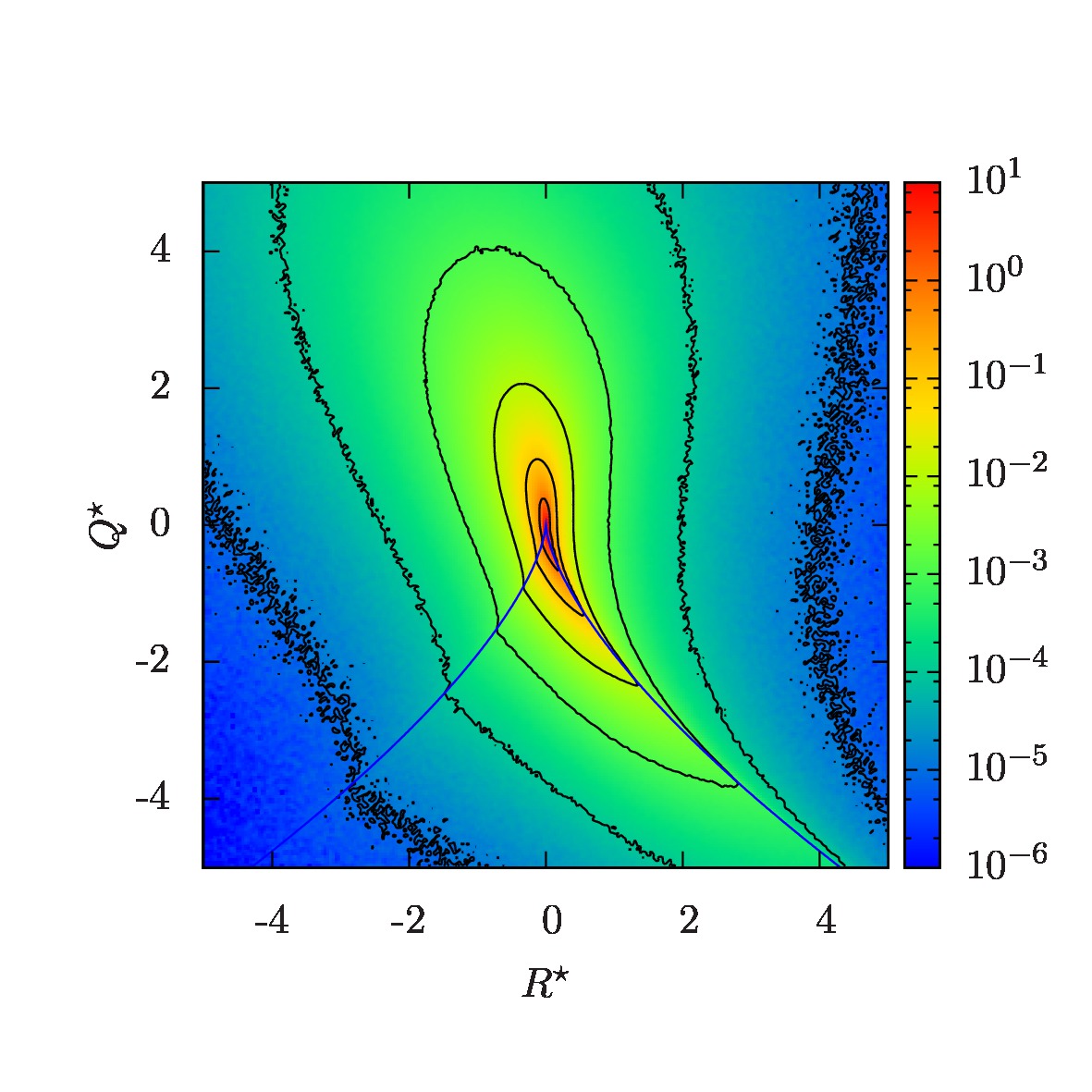}
   \includegraphics[width=0.49\textwidth]{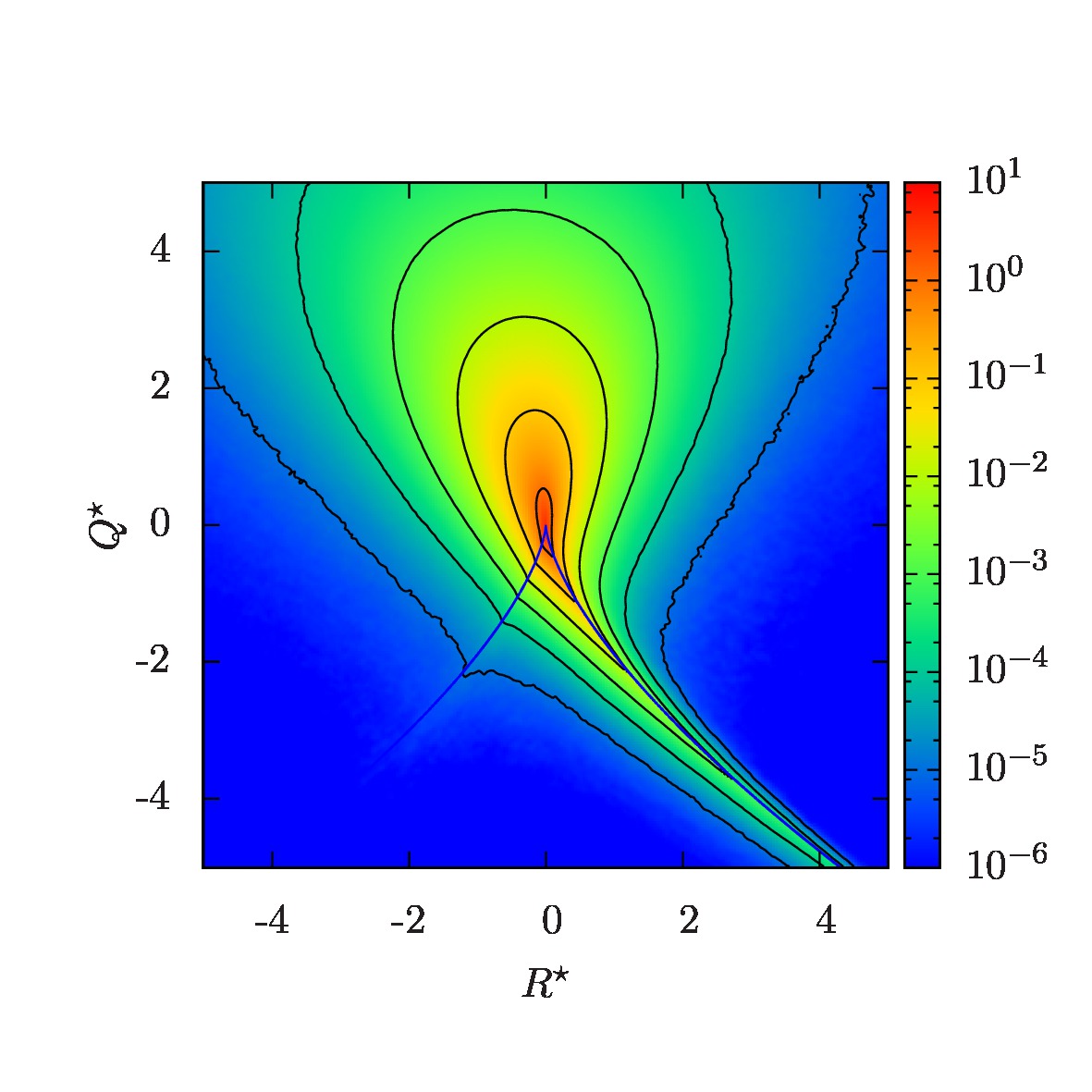}
\end{center}
\caption{Joint PDFs of $R^{\star}$ and $Q^{\star}$. Left: DNS data, right: SDE model. While the main features are captured accurately by the model, it overestimates the rotational and underestimates the straining regions in the $RQ$-plane.} \label{fig:rq_pdf}
\end{figure}

\begin{figure}
\begin{center}
  \includegraphics[width=0.49\textwidth]{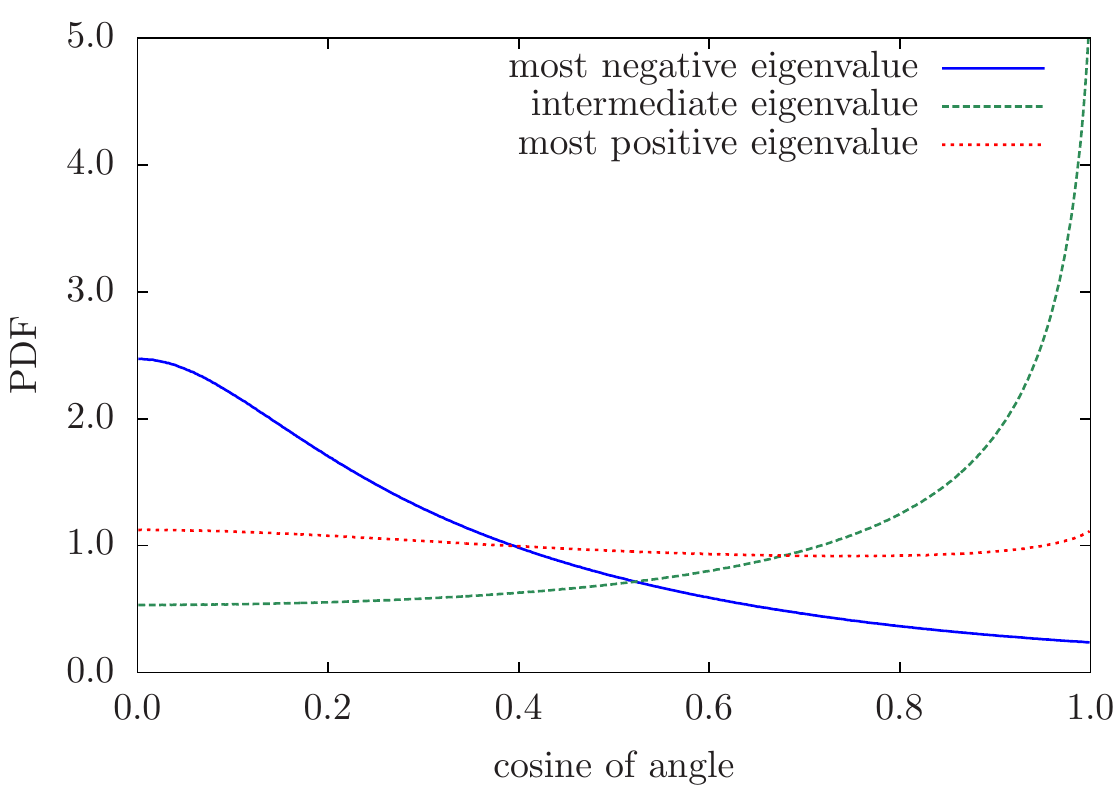}
  \includegraphics[width=0.49\textwidth]{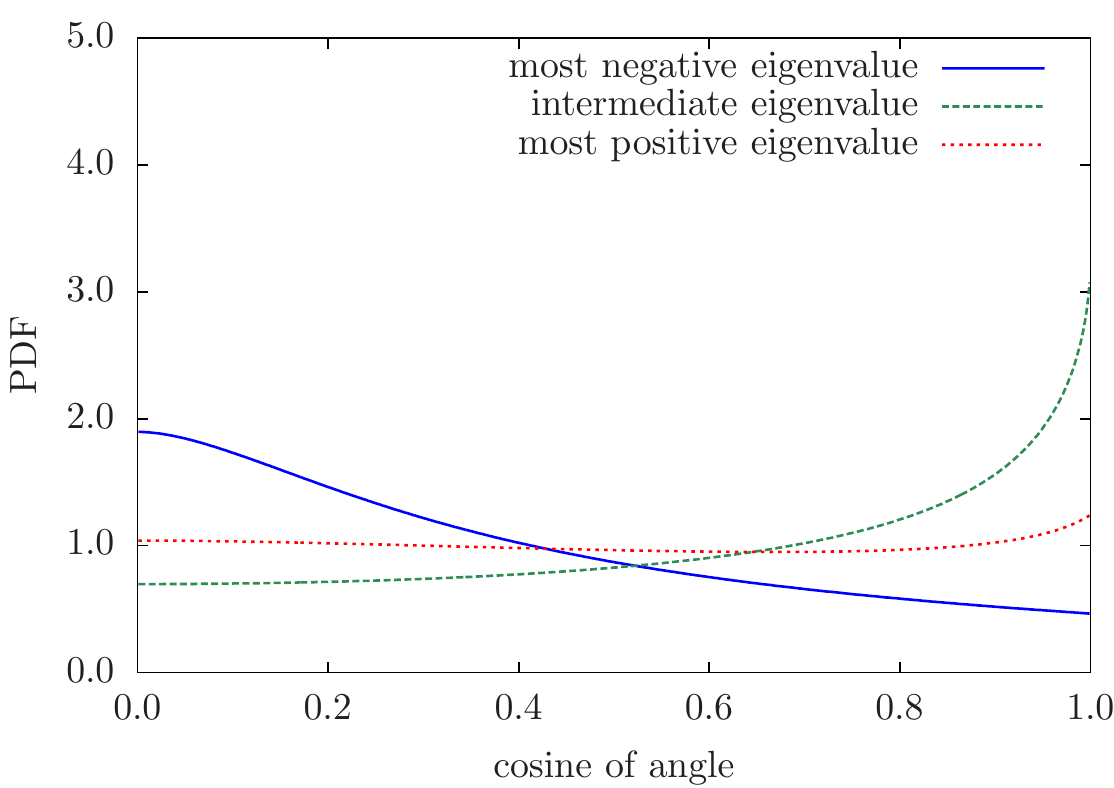}
\end{center}
\caption{PDFs of angle cosines $\widehat{\bs \omega} \cdot \bs s_i$, where $\bs s_i$ denotes the eigenvectors of the rate-of-strain tensor associated to the three eigenvalues. Left: DNS data, right: SDE model. The geometric trends of the DNS results are captured accurately by the model, although differences in the amplitudes of the PDFs can be seen.} \label{fig:alignment_pdf}
\end{figure}

Next we investigate the statistics of the model by using it in the context of a SDE. This is denoted as the `enhanced Gaussian closure SDE' model and is based on \eqref{eq:closurefullmodelnondimensionalized}. Although rooted in a Gaussian assumption for the unclosed term, this SDE will naturally produce non-Gaussian statistics due to the joint nonlinear effects of local self-amplification and non-local pressure Hessian contributions. 
The SDE of the enhanced Gaussian closure is implemented with a fourth-order Runge-Kutta method for the deterministic part, combined with an Euler method for the noise term. The details of the noise term implementation are given in appendix \ref{app:modelimplementation}.

For the numerical solution we choose the values obtained from the DNS data, $\alpha=-0.61$, $\beta=-0.65$, $\gamma=0.14$ and $\delta_{\star}=-0.15$. To determine the amplitude of the force term, $A_F$ (cf. appendix \ref{app:modelimplementation}), we note that the model should fulfil $\left\langle \mathrm{Tr}\left( \mat S_{\star}^2 \right) \right\rangle = 1/2$, which is an immediate consequence of the definition of the mean rate of kinetic energy dissipation and the non-dimensionalization leading to \eqref{eq:closurefullmodelnondimensionalized}. We ran a number of numerical tests to choose the forcing amplitude such that this constraint is fulfilled in good approximation; figure \ref{fig:force_amplitude} shows $\left\langle \mathrm{Tr}\left( \mat S_{\star}^2 \right) \right\rangle$ as a function of the forcing amplitude $A_F$.  The resulting plot evidences a strong dependence of the strain-rate variance resulting from the model as function of the forcing strength.  Still, clearly it can be seen that  choosing $A_F=0.13$ leads to $\left\langle \mathrm{Tr}\left( \mat S_{\star}^2 \right) \right\rangle \approx 1/2$. Therefore, in the numerical solutions presented below, we set $A_F=0.13$. For the numerical integration, the time step is set to $10^{-3}$. For the statistical evaluation, we draw $10^4$ initial conditions from a Gaussian ensemble (that also fulfills $\left\langle \mathrm{Tr}\left( \mat S_{\star}^2 \right) \right\rangle \approx 1/2$) and let them evolve for $5 \times 10^6$ time steps. After this initial transient, we evolve the SDE for another $5 \times 10^6$ time steps during which statistics are gathered. We have checked that all presented results are statistically well converged. However, we also noted that higher- (e.g. fourth-)order moments did not fully converge, possibly due to some rare trajectories visiting far-out regions of sample space.

Figure \ref{fig:rq_pdf} shows the PDFs of $R$ and $Q$ both from DNS and the enhanced Gaussian closure SDE. The PDFs share qualitative similarities like the characteristic tear-drop shape related to the non-vanishing enstrophy production and strain skewness. The model PDF, however, overestimates the rotational regions in the $RQ$-plane, while the straining regions, especially around the left branch of the Vieillefosse line, are underestimated. This results in the fact that the model does not respect the Betchov relations \citep{betchov56jfm} $\left\langle \mathrm{Tr}\left( \mat S^2 \right) \right\rangle=\left\langle \bs \omega^2 \right\rangle/2$ and $-\left\langle \mathrm{Tr}\left( \mat S^3 \right) \right\rangle=3\left\langle \omega_i S_{ij} \omega_j \right\rangle/4$. This discrepancy is plausible from the model pressure Hessian, which tends to suppress the statistical evolution along the right Vieillefosse tail, as well as the linearity of the diffusive term, which does not capture the details observed in the DNS. We also would like to note that Betchov's relations represent constraints of averages over fields, which are inherently difficult to incorporate into single-point models.

Focusing on geometric features of the enhanced Gaussian closure SDE, the alignment PDFs of the vorticity vector with the eigenvectors of the rate-of-strain tensor are presented in figure \ref{fig:alignment_pdf}. Results are shown for both the DNS data as well as the enhanced Gaussian closure SDE. As can be inferred from this figure, the geometrical trends observed in the DNS data are captured quite accurately by the model. While the qualitative behaviour of the alignment with the most negative and intermediate eigendirections are captured satisfactorily, differences in amplitude of the PDF are observed. 

\begin{figure}
\begin{center}
  \includegraphics[width=0.65\textwidth]{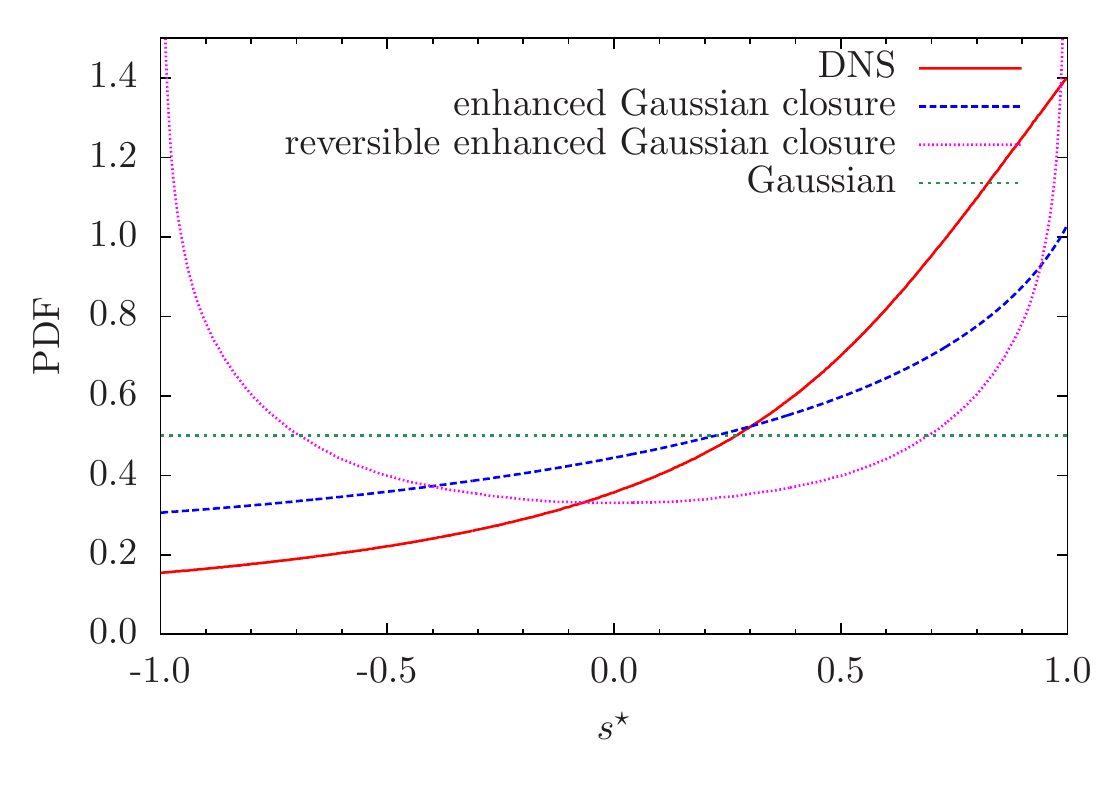}
\end{center}
\caption{PDFs of $s^{\star}=-\sqrt{6} \mathrm{Tr}\left( \mat S^3 \right) / \mathrm{Tr}\left( \mat S^2 \right)^{3/2}$ for DNS, the enhanced Gaussian closure as well as its reversible version. The flat PDF of a Gaussian random field is shown for reference.} \label{fig:sstar}
\end{figure}

The topology of the velocity gradient tensor can be conveniently described using the parameter $s^{\star}=-\sqrt{6} \mathrm{Tr}\left( \mat S^3 \right) / \mathrm{Tr}\left( \mat S^2 \right)^{3/2}$,  introduced by \citet{lund94pof}. The limiting case $s^{\star}=-1$ corresponds to a state of axisymmetric contraction, whereas $s^{\star}=1$ corresponds to axisymmetric expansion. Figure \ref{fig:sstar} shows its PDF for the DNS data and the enhanced Gaussian closure SDE results. For the DNS the PDF is a strictly increasing function showing that the preferred state of strain is axisymmetric expansion \citep{lund94pof}. This feature is qualitatively reproduced by the model, it however overestimates states of axisymmetric contraction and underestimates states of axisymmetric expansion somewhat. For a purely Gaussian field (without strain skewness, etc.), the corresponding PDF is flat.  

These comparisons show that the enhanced Gaussian closure is capable of reproducing some important qualitative features of homogeneous isotropic turbulence. For better quantitative agreement, however, more accurate modelling of the unclosed terms will be necessary. Specifically, the coefficients $\alpha$ to $\delta$ need to be generalized to depend on the invariants of the velocity gradient tensor. This, however, falls beyond the scope of the present work.

\subsection{Time-reversal symmetry}

\begin{figure}
\begin{center}
  \begin{minipage}{0.48\textwidth}
    \includegraphics[width=1.0\textwidth]{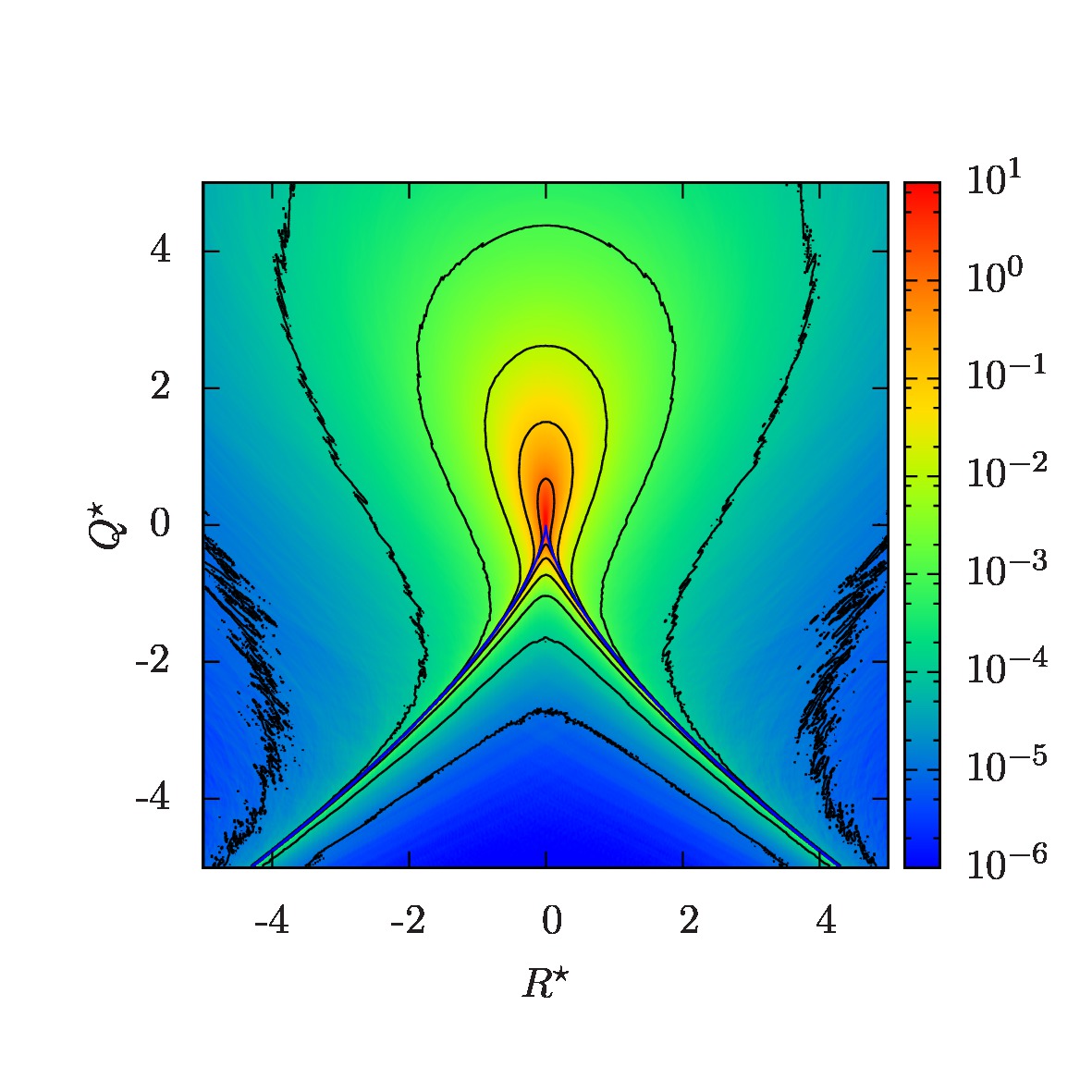}
  \end{minipage}
    \begin{minipage}{0.50\textwidth}
    \includegraphics[width=1.0\textwidth]{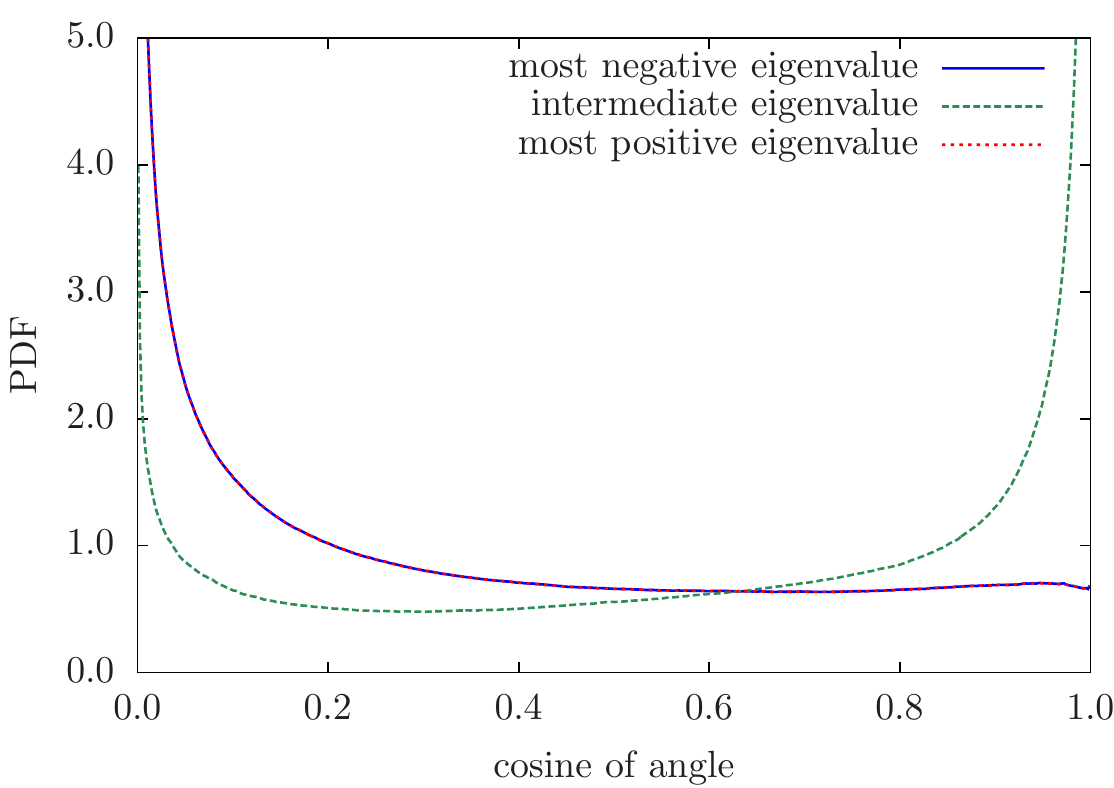}
  \end{minipage}
\end{center}
\caption{Left: $RQ$-PDF for the time-reversible enhanced Gaussian closure ODE. The $RQ$-PDF is symmetric with respect to $R \rightarrow -R$ which implies vanishing strain skewness and enstrophy production. Right: alignment PDFs. Compared with the irreversible model (see figure \ref{fig:alignment_pdf}), the alignment properties change significantly.} \label{fig:timereversal}
\end{figure}

One notable feature of the evolution equation \eqref{eq:closurefullmodel} is the time-reversal symmetry in the undamped ($\delta=0$) and unforced ($\mathrm{d}\mat F=0$) case. We have seen that the non-local pressure Hessian contributions are able to prevent a singularity for certain parameter choices, thus leading to stationary statistics. Considering the undamped and unforced case, this has the interesting consequence that all odd-order moments of the velocity gradient tensor vanish, as these quantities change sign under time reversal. This implies that the restricted Euler part of the dynamics combined with the stabilizing non-local pressure Hessian contributions considered on its own yields vanishing strain skewness, $\big\langle \mathrm{Tr}\left( \mat S^3 \right) \big\rangle=0$, as well as vanishing enstrophy production, $\big\langle \omega_i S_{ij} \omega_j \big\rangle=0$. We note in passing that, although the restricted Euler model displays the time-reversal symmetry, too, the argument does not hold in this case: The restricted Euler model does not produce stationary statistics.

These considerations can be confirmed with numerical results from the enhanced Gaussian closure ordinary differential equation (ODE; since it has no forcing, it reduces to an ODE). The PDF of $Q$ and $R$ for this case is displayed in figure \ref{fig:timereversal}. As expected, the PDF is fully symmetric with respect to the transformation $R \rightarrow -R$. Along with the $RQ$-PDF, also the PDF of vorticity alignment with the principal strain axes is shown. Interestingly, the time-reversible ODE has dramatically different alignment properties. For example, the PDFs of vorticity alignment with the most negative and most positive strain-rate eigendirection now collapse. This can be understood from the fact that the rate-of-strain tensor eigenvalues change sign under time reversal, i.e. the most positive and the most negative one change their role. As the statistics has to be invariant with respect to that transformation, the alignment PDFs of the vorticity with the most positive and most negative eigenvalue then have to collapse.

Also the PDF of $s^{\star}$ has been evaluated for the reversible case (see figure \ref{fig:sstar}). It is fully symmetric with respect to the transformation $s^{\star} \rightarrow -s^{\star}$, which also can be directly inferred by the behaviour of this quantity under time reversal.

Including linear damping and stochastic forcing, which both break this symmetry, then produces skewed statistics, non-vanishing enstrophy production and the familiar strain and alignment properties. These effects occur due to the combination of time-reversal symmetry preserving and breaking terms. Consequently, some of the essential statistical properties of small-scale turbulence cannot exclusively be associated to the closure of the pressure Hessian alone, but depend on its interplay with the dissipative and energy-injecting terms.

\section{Conclusions}

We have evaluated the effects of non-local pressure Hessian contributions and viscous diffusion in the framework of a statistical evolution equation for the velocity gradient tensor under the assumption of Gaussian incompressible velocity fields. 

In this scenario, the viscous term is obtained as a linear damping term, where the coefficient is Reynolds-number dependent through its dependence on the velocity autocorrelation function. 

The non-local contributions to the pressure Hessian are found to be a combination of quadratic, traceless and symmetric expressions of the rate-of-strain and the rate-of-rotation tensors. Two of these terms modify the original restricted Euler model with coefficients that are independent of the Reynolds number. In addition, the Gaussian closure yields a term which induces a rotation of the eigenframe of the rate-of-strain tensor, whose coefficient depends on the details of the velocity two-point correlations.

The simplicity of the Gaussian closure allowed to discuss the different dynamical effects like stretching and tilting of the vorticity vector and rate-of-strain eigenvalues and eigenvectors, revealing how the various non-local contributions of the pressure Hessian help to attenuate the occurrence of the singularity. 

The closed dynamical system has then been investigated numerically, showing that the coefficients obtained in the Gaussian approximation are insufficient to prevent the singularity caused by the restricted Euler part. Maintaining the overall structure of the Gaussian approximation, physically more realistic values for coefficients were obtained using fitting to DNS data, in a mean-field approach. The pressure Hessian and diffusive term in the enhanced Gaussian closure were shown to yield improved qualitative agreement with the DNS results.

The same applies to the statistical properties of the enhanced Gaussian closure SDE, which qualitatively captures main features such as strain skewness, enstrophy production and alignment of the vorticity with the strain eigenvectors. While the enhanced closure leads to a non-divergent time evolution, some quantitative differences to DNS were observed. A study of the closure in the time-reversible case of undamped, unforced dynamics showed some interesting results that pointed to the subtle interactions between non-local pressure contributions and the dissipative term that must be taking place in real turbulence.  

With respect to possible generalizations it is interesting to note that \eqref{eq:nonlocalpressurehessianmaintext} already represents the most general structure of the non-local pressure Hessian, which is symmetric, traceless and dimensionally consistent. In general, however, the coefficients may depend on non-dimensional combinations of the five invariants of the velocity gradient tensor. In particular, the discrepancies of the closures compared with the DNS results make coefficients that depend on, for example, strain skewness and enstrophy production plausible, which will be the topic of future research.

\section*{Acknowledgments}
We thank Gregory L.~Eyink, Laurent Chevillard, Anton Daitche and Perry Johnson for insightful discussions. Cristian C.~Lalescu is gratefully acknowledged for help with the JHU database. MW acknowledges support from the DFG under project WI 3544/2-1. CM is thankful for funding from the National Science Foundation (grants CBET 1033942 and CDI, CMMI-0941530). 

\appendix
\section{PDF equation for the velocity gradient}\label{app:pdfeqderivation}
\subsection{Derivation of the deterministic part of the PDF equation}\label{app:pdfeqderivationliouville}
This appendix is devoted to derive the evolution equation for the velocity gradient PDF and discuss the different formulations of the closure problem. As usual, the PDF is conveniently introduced as an ensemble average over a fine-grained PDF \citep{lundgren67pof,pope00book} applied to the velocity gradient:
\begin{equation}\label{eq:defpdf}
  f(\A; \bs x,t) = \big\langle \delta\left( \mat A(\bs x,t) - \A \right) \big\rangle \, .
\end{equation}
Here, $\A$ denotes the sample-space variable corresponding to the velocity gradient tensor field at position $\bs x$ and time $t$, and $f$ is a probability density with respect to $\A$ and a function with respect to $\bs x$ and $t$. We first consider the unforced case and postpone the rather technical discussion of how to include a stochastic large-scale forcing. To obtain an evolution equation for the probability density, we take the partial derivative of \eqref{eq:defpdf} with respect to time and use the chain rule and a simple change of variables:
\begin{subequations}
\begin{align}
  \frac{\partial}{\partial t} f(\A; \bs x,t) &= \left\langle \frac{\partial}{\partial t} \delta\left( \mat A(\bs x,t) - \A \right) \right\rangle \\
 &= -\frac{\partial}{\partial \Aij}\left\langle \left[\frac{\partial}{\partial t}A_{ij}(\bs x,t)\right] \delta\left( \mat A(\bs x,t) - \A \right) \right\rangle \, .
\end{align}
\end{subequations}
Here and throughout the paper summation over double indices is implied. For the time derivative of the velocity gradient field we now can substitute the dynamical equation \eqref{eq:velgradevolution} and obtain
\begin{subequations}
\begin{align}
    &\frac{\partial}{\partial t} f(\A; \bs x,t) =-\frac{\partial}{\partial \Aij}\left\langle \left[\frac{\partial}{\partial t}A_{ij}(\bs x,t)\right] \delta\left( \mat A(\bs x,t) - \A \right) \right\rangle \\
    &=-\frac{\partial}{\partial \Aij}\left\langle \left[-u_k\frac{\partial}{\partial x_k} A_{ij} -\left(A_{ik}A_{kj} - \frac{1}{3}A_{lk}A_{kl} \delta_{ij} \right) -\widetilde H_{ij}  + \nu\Delta A_{ij}\right] \delta\left( \mat A - \A \right) \right\rangle \, .
\end{align}
\end{subequations}
For brevity we have omitted the space-time dependence of all fields in the second line. This expression already shows in what way the closure problem enters this framework, namely in terms of joint averages of the various fields with the fine-grained distribution. The advective term can be treated further yielding
\begin{equation}
  -\frac{\partial}{\partial \Aij}\left\langle \left[-u_k\frac{\partial}{\partial x_k} A_{ij} \right] \delta\left( \mat A - \A \right) \right\rangle =  -\frac{\partial}{\partial x_k}\big\langle u_k \, \delta\left( \mat A - \A \right) \big\rangle \, .
\end{equation}
Here we have used again the chain rule and incompressibility of the velocity field. For homogenous flow the average of the velocity field and the delta distribution is independent of the spatial variable and hence the derivative vanishes. That means for homogeneous turbulence the advective term is gone.

Next we consider the term which is local in $\mat A$. Using the sifting property of the delta distribution, $\mat A(\bs x,t) \delta(\mat A(\bs x,t)-\A)=\A \, \delta(\mat A(\bs x,t)-\A)$, we obtain
\begin{align}
  &\left\langle \left[ A_{ik}(\bs x,t)A_{kj}(\bs x,t) - \frac{1}{3}A_{lk}(\bs x,t)A_{kl}(\bs x,t) \delta_{ij} \right] \delta\left( \mat A(\bs x,t) - \A \right) \right\rangle \nonumber \\
  &= \left[ \Aik\Akj - \frac{1}{3}\Alk\Akl \delta_{ij} \right] f(\A;\bs x,t) \, .
\end{align}
Again we have made use of the fact that sample-space variables can be pulled out of the average. The main point here is that the self-amplification term along with the isotropic part of the pressure Hessian are closed and do not need to be modeled.

For the pressure Hessian and the viscous term two options are available, namely expressing them with the help of conditional averages or in terms of multipoint statistics. We start with the former option which leads to
\begin{align}
  \big\langle \widetilde H_{ij}(\bs x,t) \delta\left( \mat A(\bs x,t) - \A \right) \big\rangle &= \big\langle \widetilde H_{ij}(\bs x,t) \big | \A \big\rangle f(\A;\bs x,t) \label{eq:conditionalpressurehessian} \\
  \big\langle \left[ \nu \Delta \mat A(\bs x,t) \right] \delta\left( \mat A(\bs x,t) - \A \right) \big\rangle &= \big\langle \nu \Delta \mat A(\bs x,t) \big | \A \big\rangle f(\A;\bs x,t) \label{eq:conditionallaplacian} \, .  
\end{align}
By introduction of conditional averages the PDF can be isolated at the price of introducing unknown functions. To close the expressions tensorial functions depending on a tensorial argument have to be modeled. The relation to the two-point statistics will be discussed later below.

\subsection{Inclusion of a stochastic force}\label{app:pdfeqderivationforcing}
To conclude the derivation of the PDF equation we consider the forcing term. If a deterministic forcing term is considered, it also can be treated with the help of conditional averages. We here, however, consider a stochastic large-scale forcing, which eventually allows us to make direct contact with earlier phenomenological models. If the Reynolds number is sufficiently high, the velocity gradient statistics should be independent of the particular choice of large-scale forcing, making the choice of this forcing not a particularly strong restriction.

The inclusion of a stochastic forcing term turns the Navier-Stokes equation to a stochastic partial differential equation. The forcing term is specified as a homogeneous isotropic Gaussian white-in-time random force $\bs F^u(\bs x,t)$  added as an additional acceleration term with the properties
\begin{equation}
  \langle \bs F^u(\bs x,t) \rangle = \bs 0 \quad \mathrm{and} \quad \langle F^u_i(\bs x,t) F^u_j(\bs x',t') \rangle = Q^u_{ij}(\bs r) \, \delta(t-t')
\end{equation}
where $\bs r=\bs x-\bs x'$.
Isotropy and solenoidality require the structure of its covariance to be
\begin{equation}
  Q^u_{ij}(\bs r) = \sigma_F^2 \left[  f_F(r) \, \delta_{ij} + \frac{1}{2} r f_F'(r) \left[ \delta_{ij} - \hat r_i \hat r_j \right] \right]
\end{equation}
where $\sigma_F^2$ denotes the forcing variance and $f_F$ denotes the longitudinal forcing autocorrelation function.

We are interested in how the stochastic forcing for the velocity field translates to the velocity gradient field. For the velocity gradient dynamics the gradient of this random force has to be added, as indicated in \eqref{eq:velgradevolution}, which then has the properties
\begin{equation}
  \langle F_{ij}(\bs x,t) \rangle = 0 \quad \mathrm{and} \quad \langle F_{ik}(\bs x,t) F_{jl}(\bs x',t') \rangle = Q_{ijkl}(\bs r) \, \delta(t-t') \, .
\end{equation}
The relations specifying the structure of the forcing resemble the results on the general structure of the velocity gradient tensor covariance discussed below in appendix \ref{app:velgradcov}, so that many of the technical results can be used here. The relation between the covariances of the random forcing for the velocity field and the velocity gradient field analogously reads
\begin{equation}
  Q_{ijkl}(\bs r) = -\frac{\partial}{\partial r_k}\frac{\partial}{\partial r_l} Q^u_{ij}(\bs r) \, .
\end{equation}
For the following we will especially need the case of $\bs r=\bs 0$, for which
\begin{equation}\label{eq:forcingcovariance}
  Q_{ijkl}(\bs 0) = -\sigma_F^2 f_F''(0) \, \left[2 \delta_{ij}\delta_{kl} -\frac{1}{2} \delta_{ik}\delta_{jl} -\frac{1}{2} \delta_{il}\delta_{jk} \right] \, .
\end{equation}
To learn how the stochastic forcing carries through to our statistical description, we adapt the presentation by \citet{haken04book} for the derivation of the Fokker-Planck equation starting from the Langevin equation and generalize it to partial differential equations. Specifically, we consider how the PDF evolves over a short time interval $\Delta t$. To this end we introduce the short-hand notation
\begin{align}
\Delta f &= f(\A;\bs x,t+\Delta t)-f(\A;\bs x,t) \quad \mathrm{and} \\
\Delta \mat A &= \mat A(\bs x,t+\Delta t)-\mat A(\bs x,t)
\end{align}
and consider furthermore
\begin{equation}\label{eq:deltaa}
 \Delta \mat A = \left[ -\bs u\cdot\nabla_{\! \bs x} \mat A - \mat A^2 - \mat H + \nu \Delta_{\bs x} \mat A\right]\Delta t + \Delta \mat F + {\cal O}\left(\Delta t^2 \right) \, .
\end{equation}
Note that, for the present discussion, $\Delta$ refers to an increment rather than the Laplacian, which here is denoted as $\Delta_{\bs x}$ to avoid ambiguities. The notation $\Delta \mat F$ for the force increment indicates that we are dealing with a stochastic force which is not differentiable in time. For the case that $\Delta t$ is \textit{infinitesimally} small, this resembles the notation of stochastic calculus, but for the moment considering a small, but finite $\Delta t$ is sufficient. Expanding the evolution of the PDF up to second order we obtain
\begin{equation}
  \Delta f = -\frac{\partial}{\partial \Aij}\big\langle \delta(\mat A - \A) \, \Delta A_{ij}\big\rangle + \frac{1}{2} \frac{\partial}{\partial \Aik}\frac{\partial}{\partial \Ajl} \big\langle \delta(\mat A - \A) \, \Delta A_{ik} \Delta A_{jl} \big\rangle + {\cal O}\left( \Delta \mat A^3 \right)  \, .\label{eq:PDFexpansionc} 
\end{equation}

Next, \eqref{eq:deltaa} is inserted into this expression, and the individual terms are evaluated with respect to their dependence on the time increment $\Delta t$. Only terms linear in $\Delta t$ are of interest, because we finally want to evaluate $\lim_{\Delta t \rightarrow 0} \Delta f/\Delta t$.

The first term in \eqref{eq:PDFexpansionc} corresponds to the deterministic contributions already discussed, plus a term involving the joint average of the force increment with the delta distribution. Because the force increment contains only contributions arising after the time $t$, the fine-grained PDF and the force increment are statistically independent, thus
\begin{equation}
  \big\langle \delta\left( \mat A - \A \right) \Delta F_{ij} \big\rangle = \big\langle \delta\left( \mat A - \A \right) \big\rangle \big\langle \Delta F_{ij} \big\rangle = 0 \, .
\end{equation}
The last equality is due to the fact that we are considering stochastic forces with zero mean. Consequently the forcing term gives no first-order contribution. This is also the reason for why we could first consider the unforced case without loss of generality.

Owing to the quadratic dependence of the second term on $\Delta \mat A$, it contains contributions proportional to $\Delta t^2$, $\Delta t \, \Delta \mat F$ and $\Delta \mat F^2$. The terms proportional to $\Delta t^2$ vanish in the limit eventually taken, and so do the terms proportional to $\Delta t \Delta \mat F$ because of the  above argument of statistical independence and  vanishing mean forcing. The only remaining term can be treated according to
\begin{subequations}
\begin{align}
   \big\langle \delta(\mat A - \A) \, \Delta F_{ik} \Delta F_{jl} \big\rangle &= \big\langle \delta(\mat A - \A) \big\rangle \big\langle \Delta F_{ik} \Delta F_{jl} \big\rangle \\
  &= \big\langle \delta(\mat A - \A) \big\rangle \, Q_{ijkl}(\bs 0) \Delta t + {\cal O}\left(\Delta t^2 \right) \, .
\end{align}
\end{subequations}
For the first equality we have again used the argument of statistical independence. To see that $\big\langle \Delta F_{ik} \Delta F_{jl} \big\rangle$ is linear in $\Delta t$, one needs to recall that the force is specified as delta-correlated in time which cancels one of the integrations necessary to evaluate the finite increment.

By evaluating the limit $\lim_{\Delta t \rightarrow 0} \Delta f/\Delta t$ and combining the results of this and the preceding section, we arrive at the PDF equation for the velocity gradient tensor in homogeneous turbulence:
\begin{align}
  \frac{\partial}{\partial t} f(\A;t) = &-\frac{\partial}{\partial \Aij} \left(\left[ -\left(\Aik\Akj-\frac{1}{3}\mathrm{Tr}\left( \A^2 \right) \delta_{ij}\right) - \big\langle \widetilde H_{ij} \big | \A \big\rangle + \big\langle \nu \Delta_{\bs x} A_{ij} \big | \A \big\rangle \right] f(\A;t)\right) \nonumber \\
  &+ \frac{1}{2} Q_{ijkl}(\bs 0)\frac{\partial}{\partial \Aik}\frac{\partial}{\partial \Ajl} f(\A;t) \, .
\end{align}

\subsection{The relation to multipoint statistics}\label{app:relationtomultipoint}

Instead of treating the unclosed terms arising in the derivation of the PDF equation with the help of conditional averages as done in \eqref{eq:conditionalpressurehessian} and \eqref{eq:conditionallaplacian}, they can also be expressed in terms of two-point statistics. In the following subscripts on the position vectors and sample-space variables will be used to discriminate the two spatial points, and $f_1$ and $f_2$ will be used for the one- and two-point PDFs, respectively. We can evaluate the viscous term according to \citep{lundgren67pof}
\begin{subequations}
\begin{align}
  &\big\langle \left[ \nu \Delta_{\bs x_1} \mat A(\bs x_1,t) \right] \delta\left( \mat A(\bs x_1,t) - \A_1 \right) \big\rangle \nonumber \\
  &= \lim_{|\bs x_2-\bs x_1|\rightarrow 0}   \big\langle \left[ \nu \Delta_{\bs x_2} \mat A(\bs x_2,t) \right] \delta\left( \mat A(\bs x_1,t) - \A_1 \right) \big\rangle \\
  &=\lim_{|\bs x_2-\bs x_1|\rightarrow 0} \nu \Delta_{\bs x_2}  \big\langle \mat A(\bs x_2,t) \big| \A_1 \big\rangle f_1(\A_1;\bs x_1,t ) \, .
\end{align}
\end{subequations}
Together with \eqref{eq:conditionallaplacian} this leads to the result
\begin{equation}
  \big\langle \nu \Delta_{\bs x_1} \mat A(\bs x_1,t) \big | \A_1 \big\rangle = \lim_{|\bs x_2-\bs x_1|\rightarrow 0} \nu \Delta_{\bs x_2}  \big\langle \mat A(\bs x_2,t) \big| \A_1 \big\rangle \, .
\end{equation}
For homogeneous turbulence, the conditional average is a function of the distance vector $\bs r=\bs x_2-\bs x_1$ only, such that we can also write
\begin{equation}
\big\langle \nu \Delta_{\bs x_1} \mat A(\bs x_1,t) \big | \A_1 \big\rangle = \lim_{r\rightarrow 0} \nu \Delta_{\bs r}  \big\langle \mat A(\bs x_2,t) \big| \A_1 \big\rangle \, .
\end{equation}
For the non-local contributions to the pressure Hessian we have to make use of the Poisson relation \eqref{eq:nonlocalpressurehessian} and also of the identity
\begin{equation}
  \int \! \mathrm{d}\A_2 \, \delta(\mat A(\bs x_2, t) - \A_2) = 1 \, .
\end{equation}
First we consider
\begin{subequations}
\begin{align}
  &\big\langle \mathrm{Tr}\left( \mat A(\bs x_2,t)^2 \right) \delta\left( \mat A(\bs x_1,t) - \A_1 \right) \big\rangle \nonumber \\
  &= \int \! \mathrm{d}\A_2 \big\langle \mathrm{Tr}\left( \mat A(\bs x_2,t)^2 \right) \delta\left( \mat A(\bs x_2,t) - \A_2 \right) \delta\left( \mat A(\bs x_1,t) - \A_1 \right) \big\rangle\\
  &= \int \! \mathrm{d}\A_2 \, \mathrm{Tr}\left( \A_2^2 \right) \, f_2(\A_1,\A_2;\bs x_1,\bs x_2,t) \\
  &= \big\langle \mathrm{Tr}\left( \mat A(\bs x_2,t)^2 \right) \big| \A_1 \big\rangle f_1(\A_1;\bs x_1,t) \, .
\end{align}
\end{subequations}
With this relation the non-local pressure Hessian can be expressed as
\begin{subequations}
\begin{align}
  &\big\langle \widetilde H_{ij}(\bs x_1,t) \delta\left( \mat A(\bs x_1,t) - \A_1 \right) \big\rangle \nonumber \\
  &=-\frac{1}{4\pi} \! \int_{\mathrm{P.V.}} \!\!\!\!\, \mathrm{d}\bs x_2 \left[ \frac{\delta_{ij}}{|\bs x_2-\bs x_1|^3}-3\frac{(\bs x_2-\bs x_1)_i (\bs x_2-\bs x_1)_j}{|\bs x_2-\bs x_1|^5}  \right]  \big\langle \mathrm{Tr}\left( \mat A(\bs x_2,t)^2 \right) \delta\left( \mat A(\bs x_1,t) - \A_1 \right) \big\rangle \\
 &=-\frac{1}{4\pi} \! \int_{\mathrm{P.V.}} \!\!\!\!\, \mathrm{d}\bs x_2 \left[ \frac{\delta_{ij}}{|\bs x_2-\bs x_1|^3}-3\frac{(\bs x_2-\bs x_1)_i (\bs x_2-\bs x_1)_j}{|\bs x_2-\bs x_1|^5}  \right] \big\langle \mathrm{Tr}\left(\mat A(\bs x_2,t)^2\right) \big | \A_1 \big\rangle f_1(\A_1;\bs x_1,t) . 
\end{align}
\end{subequations}

Together with \eqref{eq:conditionalpressurehessian} this leads to the result
\begin{equation}
\big\langle \widetilde H_{ij}(\bs x_1,t) \big | \A_1 \big\rangle = \frac{1}{2\pi} \! \int_{\mathrm{P.V.}} \!\!\!\!\, \mathrm{d}\bs x_2 \left[ \frac{\delta_{ij}}{|\bs x_2-\bs x_1|^3}-3\frac{(\bs x_2-\bs x_1)_i (\bs x_2-\bs x_1)_j}{|\bs x_2-\bs x_1|^5}  \right] \big\langle Q(\bs x_2,t) \big | \A_1 \big\rangle \, ,
\end{equation}
and for homogeneous turbulence
\begin{equation}
\big\langle \widetilde H_{ij}(\bs x_1,t) \big | \A_1 \big\rangle = \frac{1}{2\pi} \! \int_{\mathrm{P.V.}} \!\!\!\!\, \mathrm{d}\bs r \left[ \frac{\delta_{ij}}{r^3}-3\frac{r_i r_j}{r^5}  \right] \big\langle Q(\bs x_2,t) \big | \A_1 \big\rangle \, . \label{eq:conditionalpressurehessianintermsofq}
\end{equation}

\section{Gaussian approximation}

\subsection{Velocity gradient covariance tensor}\label{app:velgradcov}

The general structure of the velocity gradient covariance tensor is obtained by evaluating the kinematic relation \eqref{eq:velgradcovariance}. We also make use of homogeneity, which implies that the covariance tensor depends on $\bs r = \bs x-\bs x'$ only, and isotropy. As a result we obtain
\begin{align}\label{eq:covariancestructure}
  R_{ijkl}(\bs r) &= a_1 \, \delta_{ij}\delta_{kl} + a_2 \, \big[ \delta_{ik}\delta_{jl} + \delta_{il}\delta_{jk} \big] + a_3 \, \delta_{ij} \, \hat{r}_k\hat{r}_l \nonumber \\
   &+ a_4 \, \big[ \delta_{ik} \, \hat{r}_j\hat{r}_l + \delta_{il} \, \hat{r}_j\hat{r}_k + \delta_{jk} \, \hat{r}_i\hat{r}_l + \delta_{jl} \, \hat{r}_i\hat{r}_k + \delta_{kl} \, \hat{r}_i\hat{r}_j \big] + a_5 \, \hat{r}_i \hat{r}_j \hat{r}_k \hat{r}_l
\end{align}
where the coefficients depend on the longitudinal velocity autocorrelation function and are given by
\begin{subequations}\begin{align}
  a_1 &= \frac{\langle \bs u^2 \rangle}{6} \left[ -3\frac{f_u'}{r} - f_u'' \right] \\
  a_2 &= \frac{\langle \bs u^2 \rangle}{6} \left[ \frac{f_u'}{r} \right] \\
  a_3 &= \frac{\langle \bs u^2 \rangle}{6} \left[ 3\frac{f_u'}{r} - 3f_u'' -r f_u'''\right] \\
  a_4 &= \frac{\langle \bs u^2 \rangle}{6} \left[ -\frac{f_u'}{r} + f_u'' \right]\\
  a_5 &= \frac{\langle \bs u^2 \rangle}{6} \left[ 3\frac{f_u'}{r} - 3f_u'' + r f_u'''\right] \, .
\end{align}\end{subequations}

Some interesting observations can be made here. Incompressibility of the velocity field implies $A_{ii}=0$. For the covariance tensor this implies $R_{ijil}=0$ and $R_{ijkj}=0$ which is readily checked with the above results. We also need to explicitly evaluate the tensor for $\bs r=\bs 0$, in which case only the coefficients $a_1$ and $a_2$ should remain. Indeed, by making use of L'Hospital's rule we obtain $a_3=a_4=a_5=0$ and
\begin{subequations}\begin{align}
  a_1(0) &= \lim_{r \rightarrow 0} \frac{\langle \bs u^2 \rangle}{6} \left[ -3\frac{f_u'}{r} - f_u'' \right] = -2\frac{\langle \bs u^2 \rangle}{3} f_u''(0) = \frac{2}{15} \frac{\varepsilon}{\nu} \\
  a_2(0) &= \lim_{r \rightarrow 0} \frac{\langle \bs u^2 \rangle}{6} \left[ \frac{f_u'}{r} \right] = \frac{\langle \bs u^2 \rangle}{6} f_u''(0)  = -\frac{1}{30} \frac{\varepsilon}{\nu} \, .
\end{align}\end{subequations} 
For these calculations we have made use of the properties $f_u'(0)=0$ and the relation (see, e.g., \cite{pope00book})
\begin{equation}
  \frac{\langle \bs u^2 \rangle}{3} f_u''(0) = -\frac{1}{15} \frac{\varepsilon}{\nu}=-\frac{2}{15} \left\langle \mathrm{Tr}\left( \mat S^2 \right) \right\rangle \, .
\end{equation}
This leads to the result
\begin{equation}
  R_{ijkl}(\bs 0) = \frac{\langle \bs u^2 \rangle}{6} f_u''(0) \, \big[-4 \delta_{ij}\delta_{kl} + \delta_{ik}\delta_{jl} + \delta_{il}\delta_{jk} \big] \, .
\end{equation}

For the Fourier transform of the single-point characteristic function to the single-point PDF of the velocity gradient tensor also the inverse of this expression is needed. We construct the inverse by considering
\begin{equation}\label{eq:inverse}
  R_{ijkl}(\bs 0)R^{-1}_{jnlp}(\bs 0)\Anp = \Aik \, .
\end{equation}
In this context it is important to note that the velocity gradient covariance tensor is singular due to solenoidality of the velocity field.  Consistently taking into account $\Aii=0$, however, still allows to introduce the above definition of an inverse. It is readily checked that for traceless matrices the result reads
\begin{equation}\label{eq:inversecovariance}
  R^{-1}_{jnlp}(\bs 0) = \frac{2}{5} \frac{1}{\langle \bs u^2 \rangle f_u''(0)}  \big[ -4 \delta_{jn}\delta_{lp} - \delta_{jp}\delta_{ln} \big] \, .
\end{equation}

\subsection{First conditional moment and conditional Laplacian}\label{app:firstmomentandlaplacian}

To explicitly obtain the conditional Laplacian \eqref{eq:conditionallaplaciantwopoint} in the Gaussian approximation, we first have to evaluate the first conditional moment in the Gaussian approximation. To this end we consider the definition of the conditional moment
\begin{equation}\label{eq:deffirstconditionalmoment}
  \big\langle \mat A(\bs x_2) \big | \A_1 \big\rangle g_1(\A_1) = \int \! \mathrm{d}\A_2 \, \A_2 \, g_2(\A_1,\A_2) \, . 
\end{equation}
Here $g_1$ and $g_2$ denote the single- and two-point Gaussian distributions, respectively. For the following calculation we especially need the characteristic function of the two-point Gaussian distribution \eqref{eq:twopointcharfun}, which is related to the two-point PDF by inverse Fourier transform:
\begin{equation}
  g_2(\A_1,\A_2) = (2\pi)^{-18}\int \! \mathrm{d}\mat \Lambda_1 \, \mathrm{d}\mat \Lambda_2 \, \phi^{\mat A}_2(\mat \Lambda_1,\mat \Lambda_2) \, \exp\left[ -\mathrm{i} \left( \Lambda_{1,kl}\Aonekl + \Lambda_{2,kl}\Atwokl \right) \right] \, .
\end{equation}
Inserting the last expression into \eqref{eq:deffirstconditionalmoment} and noticing that
\begin{equation}\label{eq:derivativetrick1}
  \Atwoij \exp\left[ -\mathrm{i} \left( \Lambda_{1,kl}\Aonekl + \Lambda_{2,kl}\Atwokl \right) \right] = \mathrm{i} \frac{\partial}{\partial \Lambda_{2,ij}} \exp\left[ -\mathrm{i} \left( \Lambda_{1,kl}\Aonekl + \Lambda_{2,kl}\Atwokl \right) \right] 
\end{equation}
we can make use of partial integration to obtain
\begin{align}\label{eq:firstconditionalmomentintermediatestep}
  &\big\langle A_{ij}(\bs x_2) \big | \A_1 \big\rangle g_1(\A_1) =\nonumber \\ 
  &-\mathrm{i}(2\pi)^{-18}\int \! \mathrm{d}\A_2 \, \mathrm{d}\mat \Lambda_1 \, \mathrm{d}\mat \Lambda_2 \, \left[\frac{\partial}{\partial \Lambda_{2,ij}}\phi^{\mat A}_2(\mat \Lambda_1,\mat \Lambda_2)\right] \, \exp\left[ -\mathrm{i} \left( \Lambda_{1,kl}\Aonekl + \Lambda_{2,kl}\Atwokl \right) \right] \, .
\end{align}
The derivative of the two-point characteristic function is readily obtained and reads
\begin{equation}\label{eq:firstderivativecharfun}
  \frac{\partial}{\partial \Lambda_{2,ij}}\phi^{\mat A}_2(\mat \Lambda_1,\mat \Lambda_2) = -\left[ R_{ikjl}(\bs r)\Lambda_{1,kl}+R_{ikjl}(\bs 0)\Lambda_{2,kl}  \right] \, \phi^{\mat A}_2(\mat \Lambda_1,\mat \Lambda_2) \, .
\end{equation}
For this result we have also made use of the fact that $R_{ijkl}$ for homogeneous isotropic flows remains unchanged under a simultaneous change of indices $i \leftrightarrow j$ and $k \leftrightarrow l$. In analogy to \eqref{eq:derivativetrick1} we have the relation
\begin{equation}\label{eq:derivativetrick2}
  \Lambda_{1,ij} \exp\left[ -\mathrm{i} \left( \Lambda_{1,kl}\Aonekl + \Lambda_{2,kl}\Atwokl \right) \right] = \mathrm{i} \frac{\partial}{\partial \Aoneij} \exp\left[ -\mathrm{i} \left( \Lambda_{1,kl}\Aonekl + \Lambda_{2,kl}\Atwokl \right) \right] \,
\end{equation}
which together with \eqref{eq:firstderivativecharfun} and \eqref{eq:firstconditionalmomentintermediatestep} can be used to obtain the relation
\begin{align}
  \big\langle A_{ij}(\bs x_2) \big | \A_1 \big\rangle g_1(\A_1) = -R_{ikjl}(\bs r)\frac{\partial}{\partial \Aonekl} g_1(\A_1) \, ,
\end{align}
where we have carried out the integration with respect to $\A_2$ and identified the one-point PDF as the inverse Fourier transform of the one-point characteristic function. The derivative of the Gaussian single-point PDF is readily obtained and yields
\begin{equation}
  \frac{\partial}{\partial \Aonekl} g_1(\A_1) = -R^{-1}_{kmln}(\bs 0)\Aonemn \, g_1(\A_1) \, ,
\end{equation}
such that we obtain the final result
\begin{equation}
  \big\langle A_{ij}(\bs x_2) \big | \A_1 \big\rangle = R_{ikjl}(\bs r)R^{-1}_{kmln}(\bs 0)\Aonemn \, .
\end{equation}
By \eqref{eq:conditionallaplaciantwopoint} we now have to evaluate the Laplacian of this expression (times the kinematic viscosity) which comes down to calculating $\lim_{r \rightarrow 0}\Delta_{\bs r} R_{ikjl}(\bs r)$. The calculation involves again a careful application of L'Hospital's rule and yields the result
\begin{equation}
  \lim_{r \rightarrow 0} \, \Delta_{\bs r} R_{ikjl}(\bs r) = \frac{7}{18} \langle \bs u^2 \rangle f_u^{(4)}(0) \big[ -4 \delta_{ik}\delta_{jl} + \delta_{ij}\delta_{kl} + \delta_{il}\delta_{jk} \big] \, .
\end{equation}
Together with the inverse \eqref{eq:inversecovariance} we obtain for the conditional Laplacian
\begin{subequations}
\begin{align}
  \big\langle \nu \Delta_{\bs x_1} A_{ij}(\bs x_1) \big | \A_1 \big\rangle &= \nu \left[ \lim_{r \rightarrow 0} \Delta_{\bs r} R_{ikjl}(\bs r)\right] R^{-1}_{kmln}(\bs 0)\Aonemn \\
  &= \nu \frac{7}{3} \, \frac{f_u^{(4)}(0)}{f_u''(0)} \, \Aoneij \, . \label{eq:gaussianconditionallaplacian}
\end{align}
\end{subequations}
The prefactor of \eqref{eq:gaussianconditionallaplacian} can also be conveniently expressed in terms of the energy spectrum function. To establish the relation, we have to express the longitudinal velocity autocorrelation function in terms of the energy spectrum function, which can be achieved by considering
\begin{equation}
  R^u_{ij}(\bs r)  = \int  \! \mathrm{d} \bs k \, \phi_{ij}(\bs k) \, \exp[\mathrm{i}\bs k\cdot \bs r]
\end{equation}
where
\begin{equation}
  \phi_{ij}(\bs k) = \frac{E(k)}{4 \pi k^2} \left( \delta_{ij} - \frac{k_i k_j}{k^2} \right)
\end{equation}
is the energy spectrum tensor and $E(k)$ the energy spectrum function. This relation can be used to evaluate
\begin{subequations}
\begin{align}
  \lim_{r\rightarrow 0} \frac{\partial^2}{\partial r_l \partial r_l} R^u_{ii}(\bs r)  &= \frac{\langle \bs u^2 \rangle}{3} 15 f_u''(0) \\
  &= -2\int \! \mathrm{d}k \, k^2 \, E(k)
\end{align}
\end{subequations}
leading to
\begin{equation}
 f_u''(0) = -\frac{2}{15} \frac{3}{\langle \bs u^2 \rangle} \int \! \mathrm{d}k \, k^2 \, E(k) \, .
\end{equation}
Similar calculations can be performed to derive the relation
\begin{equation}
 f_u^{(4)}(0) = \frac{2}{35} \frac{3}{\langle \bs u^2 \rangle} \int \! \mathrm{d}k \, k^4 \, E(k) \, .
\end{equation}
As a result, \eqref{eq:gaussianconditionallaplacian} takes the simple form
\begin{equation}\label{eq:deltaspectrum}
 \big\langle \nu \Delta_{\bs x_1} \mat A(\bs x_1) \big| \A_1 \big\rangle = -\nu  \frac{\int \! \mathrm{d}k \, k^4 \, E(k)}{\int \! \mathrm{d}k \, k^2 \, E(k)} \, \A_1 \, .
\end{equation}

\subsection{Second conditional moment and conditional pressure Hessian}\label{app:secondmomentandhessian}

To evaluate the non-local contributions to the pressure Hessian in the Gaussian approximation following \eqref{eq:conditionalpressurehessianmain}, we first need to evaluate the conditional second invariant
\begin{equation}
  \big\langle Q(\bs x_2) \big | \A_1 \big\rangle = -\frac{1}{2} \big\langle A_{mn}(\bs x_2)A_{nm}(\bs x_2) \big | \A_1 \big\rangle \, .
\end{equation}
The evaluation of the second conditional moment
\begin{equation}
  \big\langle A_{mn}(\bs x_2)A_{nm}(\bs x_2) \big | \A_1 \big\rangle g_1(\A_1) = \int \! \mathrm{d}\A_2 \, \Atwomn\Atwonm \, g_2(\A_1,\A_2)
\end{equation}
is analogous to the calculation of the first conditional moment and resembles many of the steps of the prior section. The result for the conditional second invariant eventually reads
\begin{align}\label{eq:conditionalQ}
 \big\langle Q(\bs x_2) \big| \A_1 \big\rangle = & -\frac{4}{5}\frac{1}{\langle \bs u^2 \rangle f_u''(0)} \, R_{mink}(\bs r)R_{nimk}(\bs r) -\frac{1}{5}\frac{1}{\langle \bs u^2 \rangle f_u''(0)} R_{mink}(\bs r)R_{nkmi}(\bs r) \nonumber \\
 -&\frac{2}{25} \left(\frac{1}{\langle \bs u^2 \rangle f_u''(0)} \right)^2 R_{mink}(\bs r)R_{njml}(\bs r) \left[ 4\Aoneik + \Aoneki \right]\left[ 4\Aonejl + \Aonelj \right] \, . 
\end{align}
The result is a combination of tensor contractions of the velocity gradient covariance tensor and the velocity gradient tensor. It depends on the longitudinal velocity autocorrelation function through the velocity gradient covariance tensor, which can be seen when evaluating this expression for the general structure \eqref{eq:covariancestructure} of the velocity gradient covariance tensor in homogeneous isotropic turbulence:
\begin{align}\label{eq:conditionalqcorrelationfunction}
   \big\langle Q(\bs x_2) \big| \A_1 \big\rangle =   &-\frac{1}{450 f_u''(0)^2} \bigg(20 b_1 \langle \bs u^2 \rangle f_u''(0) + b_2 \, \mathrm{Tr}\left( \A_1^2 \right) + b_3 \, \mathrm{Tr}\left( \A_1\A_1^{\mathrm{T}} \right) + b_4 \, \left(\hat{\bs r}^{\mathrm{T}} \A_1^2 \hat{\bs r}\right) \nonumber \\
  &+ b_5 \, \left(\hat{\bs r}^{\mathrm{T}} \A_1\A_1^{\mathrm{T}} \hat{\bs r}\right) + b_6 \, \left(\hat{\bs r}^{\mathrm{T}} \A_1^{\mathrm{T}}\A_1 \hat{\bs r}\right) + b_7 \, \left(\hat{\bs r}^{\mathrm{T}} \A_1 \hat{\bs r}\right)^2 \bigg)
\end{align}
with
\begin{subequations}\begin{align}
  b_1 &=  -\left(\frac{f_u'}{r}\right)^2 +2\,\frac{f_u'}{r}f_u'' +2\,f_u'\,f_u'''-f_u''^2 -2\,rf_u''\,f_u''' \\
  b_2 &= 122 \left(\frac{f_u'}{r}\right)^2 +86 \frac{f_u'}{r}f_u'' +17 f_u''^2  \\
  b_3 &= -22 \left(\frac{f_u'}{r}\right)^2 +14 \frac{f_u'}{r}f_u'' +8 f_u''^2 \\
  b_4 &= -244 \left( \frac{f_u'}{r} \right)^2 +308 \frac{f_u'}{r} f_u''  +136 f_u' f_u''' -64 f_u''^2 -16 r f_u'' f_u'''  \\
  b_5 &= 22 \left(\frac{f_u'}{r}\right)^2 -14 \frac{f_u'}{r} f_u'' +32 f_u' f_u''' - 8 f_u''^2 -2 r f_u'' f_u'''   \\
  b_6 &= 22 \left( \frac{f_u'}{r} \right)^2 +106 \frac{f_u'}{r}f_u'' +32 f_u'f_u''' -128 f_u''^2  -32 r f_u''f_u''' \\
  b_7 &= 50 \left(\frac{f_u'}{r}\right)^2  -550 \frac{f_u'}{r} f_u'' -200 f_u'f_u''' +500 f_u''^2 +50 r f_u'' f_u''' \, .
\end{align}\end{subequations}

As expected these terms turn out to be invariants composed of $\A_1$ and $\hat{\bs r}$. It is interesting to check the limiting behaviour of this expression. In the limit of $r\rightarrow\infty$, the conditional second invariant vanishes due to the decay of correlations. In the limit $r \rightarrow 0$, $b_1$ as well as $b_3$ through $b_7$ vanish, as a careful evaluation with L'Hospital's rule shows. For $b_2$, however, we obtain $b_2=225f_u''(0)^2$ in this limit, which yields the expected limiting behaviour
\begin{equation}
  \lim_{r \rightarrow 0} \big\langle Q(\bs x_2) \big| \A_1 \big\rangle = Q_1 \, .
\end{equation}
Inserting expression \eqref{eq:conditionalqcorrelationfunction} into \eqref{eq:conditionalpressurehessianmain}, one realizes that the term independent of $\A_1$ does not give any contribution. A lengthy calculation reveals that the terms quadratic in $\A_1$ can be grouped taking the form\begin{align}
  \big\langle \mat{\widetilde{H}}(\bs x_1) \big| \A_1 \big\rangle &=  \alpha \, \left( \S_1^2 - \frac{1}{3}\mathrm{Tr}\left( \S_1^2 \right) \mat I \right) \nonumber \\
                      									    &+  \beta \, \left( \W_1^2 - \frac{1}{3}\mathrm{Tr}\left( \W_1^2 \right) \mat I \right) \nonumber \\
									       		    &+  \gamma \, \left( \S_1\W_1 - \W_1\S_1  \right)
\end{align}
with
\begin{subequations}
\begin{align}
  \alpha &= -\frac{4}{105 f_u''(0)^2} \int \! \mathrm{d}r \left( 8\frac{f_u'^2}{r^3} -4 \frac{f_u'f_u''}{r^2}  -4 \frac{f_u'f_u'''}{r}  -4 \frac{f_u''^2}{r} + f_u''f_u''' \right) \\
  \beta &= -\frac{4}{125 f_u''(0)^2} \int \! \mathrm{d}r \left( 16 \frac{f_u'^2}{r^3} -12\frac{f_u'f_u''}{r^2} -4 \frac{f_u'f_u'''}{r} -4 \frac{f_u''^2}{r}- f_u''f_u''' \right) \\
  \gamma &= \frac{4}{75 f_u''(0)^2} \int \! \mathrm{d}r \left( 4\frac{f_u'f_u''}{r^2} -4\frac{f_u''^2}{r}-f_u''f_u''' \right) \, .
\end{align}
\end{subequations}
These terms can be significantly simplified by partial integration and identifying product rules, which then leads to
\begin{subequations}\begin{align}
  \alpha &= -\frac{2}{7}  \\
  \beta &= -\frac{2}{5} \\
  \gamma &=  \frac{6}{25} + \frac{16}{75 f_u''(0)^2} \int \! \mathrm{d}r \, \frac{f_u'f_u'''}{r} \label{eq:gamma} \, .
\end{align}\end{subequations}
That means, the coefficients $\alpha$ and $\beta$ become independent of the longitudinal velocity autocorrelation function, whereas this dependence remains for the term $\gamma$ through a nonlinear integral dependence.

\section{Parameter estimation from a model spectrum}\label{app:modelspectrum}

The parameters $\gamma$ and $\delta$ from the Gaussian approximation depend implicitly on the Reynolds number through the longitudinal velocity autocorrelation function or, equivalently, through the energy spectrum function. To obtain approximate values for the parameters we use a simplified model spectrum similar to those proposed by \cite{meyers08pof} or \cite{pope00book} (but unlike that proposed by \cite{meyers08pof}, without intermittency and bottleneck corrections). We use the basic form
\begin{subequations}
\begin{align}
  E(k) &= C_K\varepsilon^{2/3}k^{-5/3} \, F_{L}(k L) \, F_{\eta}(k \eta) \quad \mathrm{with} \\
  F_{L}(k L) &= \left[ \frac{c_1 k L}{\left[ (c_1 k L)^{3/2} + 1 \right]^{2/3}} \right]^{5/3+2} \quad \mathrm{and} \\
  F_{\eta}(k\eta) &= \exp\left( -c_2k\eta \right) \, .
\end{align}
\end{subequations}
Here, $C_K$ is the Kolmogorov constant, $\varepsilon$ is the dissipation rate, $L$ denotes the integral length scale, $\eta$ denotes the Kolmogorov length scale and $c_1$ and $c_2$ are non-dimensional parameters which are determined such that the length scales in the above expression are consistent with the relations $L = (2E_{\mathrm{kin}}/3)^{3/2}/\varepsilon$ and $\eta = (\nu^3/\varepsilon)^{1/4}$. The kinetic energy and the rate of energy dissipation are given by the integrals
\begin{align}
  E_{\mathrm{kin}} = &\int\!\mathrm{d}k \, E(k) \\
  \varepsilon = 2\nu&\int\!\mathrm{d}k \, k^2 \, E(k) \, . \label{eq:epsilonintegral}
\end{align}
For a given dissipation rate, the kinematic viscosity then is determined by \eqref{eq:epsilonintegral}, and the Taylor Reynolds number can be estimated as $R_{\lambda}=\sqrt{20/3}E_{\mathrm{kin}}/\sqrt{\varepsilon\nu}$. For a given energy spectrum function, the longitudinal velocity autocorrelation is given by
\begin{equation}
  f_u(r) = \frac{6}{\langle \bs u^2 \rangle} \int\!\mathrm{d}k \, E(k) \left[ \frac{\sin(kr)}{(kr)^3} - \frac{\cos(kr)}{(kr)^2} \right] \, .
\end{equation}
Also derivatives of the velocity autocorrelation function can be obtained straightforwardly, such that we can evaluate \eqref{eq:gammamaintext} to obtain a numerical estimate for $\gamma$; $\delta$ is readily obtained from a numerical evaluation of \eqref{eq:deltamaintext} in terms of the energy spectrum function.

For the present example we choose $C_K = 1.6$, $\varepsilon=1.0 \, \mathrm{m}^2/\mathrm{s}^3$ as well as $c_1L=1.0 \, \mathrm{m}$ and $c_2\eta=0.002 \, \mathrm{m}$, which results in a Reynolds number of $R_{\lambda}\approx 432$. For this model spectrum we obtain $\gamma \approx0.08$ and $\delta \, \tau_{\eta}\approx -0.65$. 

\section{Implementation of noise term in the SDE}\label{app:modelimplementation}
For the discussion of the implementation of the noise term we mainly follow \cite{chevillard08pof}, to which we refer the reader for more background information. Note that our presentation differs with respect to some coefficients (which are a matter of convention). The noise term in \eqref{eq:closurefullmodelnondimensionalized} can be written as \citep{chevillard08pof}
\begin{equation}
  \mathrm{d}F_{\star ij} = A_F D_{ijkl}\mathrm{d}W_{\star kl} \, ,
\end{equation}
where $A_F$ is the forcing amplitude and $\mathrm{d}W_{\star kl}$ in this section denotes an isotropic tensorial Wiener process specified by $\langle \mathrm{d} \mat W_{\star} \rangle = 0$ and $\langle \mathrm{d} W_{\star ik} \mathrm{d} W_{\star jl} \rangle = \delta_{ij}\delta_{kl} \mathrm{d} t_{\star}$. Here $\mat D$ is a tensor specified such that the noise term complies with the tensorial structure of the forcing applied to the velocity gradient tensor evolution equation discussed in appendix \ref{app:pdfeqderivationforcing}. We recall that this forcing is specified by  $\langle \mathrm{d} \mat F_{\star} \rangle = 0$ and $\langle \mathrm{d} F_{\star ik} \mathrm{d} F_{\star jl} \rangle = Q_{\star ijkl}(\bs 0) \mathrm{d} t_{\star}$, where the $\mat Q_{\star}=\mat Q \tau_{\eta}^3$ is the non-dimensionalized version of the force covariance tensor \eqref{eq:forcingcovariance}. Now identifying the forcing amplitude as $A_F = \big(-\sigma_F^2 f_F''(0) \tau_{\eta}^3 \big)^{1/2}$ leads to the condition
\begin{equation}
 D_{ikmn} D_{jlmn}   = 2 \delta_{ij}\delta_{kl} -\frac{1}{2} \delta_{ik}\delta_{jl} -\frac{1}{2} \delta_{il}\delta_{jk}\, .
\end{equation}
The solution to this equation has been worked out by \cite{chevillard08pof} and reads
\begin{equation}
  D_{ijkl} = \frac{1}{3} \frac{3 + \sqrt{15} }{\sqrt{10} + \sqrt{6}} \delta_{ij} \delta_{kl} -\frac{\sqrt{10}+\sqrt{6}}{4} \delta_{ik} \delta_{jl} + \frac{1}{\sqrt{10}+\sqrt{6}} \delta_{il} \delta_{jk} \, ,
\end{equation}
which concludes the specification of the stochastic forcing term.


\end{document}